\documentstyle[aps,epsfig,preprint]{revtex}

\begin{document}  

\tighten 

\title{Interrelation between the isoscalar octupole phonon 
       and the proton-neutron mixed-symmetry quadrupole phonon 
       in near spherical nuclei}

\author{Nadya A. Smirnova$^{1,2,3}$, Norbert Pietralla$^{4,5}$, 
        Takahiro~Mizusaki$^{6}$, Piet Van~Isacker$^{1}$}

\address{$^{1}$ GANIL, BP 5027, F-14076, Caen Cedex 5, France }  
\address{$^{2}$ Centre de Spectrom\'etrie Nucl\'eaire et de Spectrom\'etrie 
de Masse, F-91405, Orsay, France }
\address{$^{3}$ Institute for Nuclear Physics, Moscow State University,
119899 Moscow, Russia}    
\address{$^{4}$ Institut f\"ur Kernphysik, Universit\"at zu K\"oln, 
50937 K\"oln, Germany}  
\address{$^{5}$ Wright Nuclear Structure Laboratory, Yale University, 
New Haven, Connecticut 06520}  
\address{$^{6}$ Department of Law, Senshu University, 1-1, 
Higashimita 2-chome, Tama-ku, Kawasaki-shi, Kanagawa, 214-8580, Japan}
 
%\date{\today}

\maketitle

\vspace{1cm}

\noindent
Corresponding author: \\
Nadya A. Smirnova\\
Centre de Spectrom\'etrie Nucl\'eaire et de Spectrom\'etrie 
de Masse, \\
Bat. 104, Orsay campus, F-91405, France \\
Tel: (+33) (0)1 69 15 48 57 \\
Fax: (+33) (0)1 69 15 50 08\\
E-mail: smirnova@csnsm.in2p3.fr\\

\begin{abstract}
The interrelation between the octupole phonon and 
the low-lying proton-neutron mixed-symmetry quadrupole phonon 
in near-spherical nuclei is investigated. 
The one-phonon states decay by collective $E3$ and $E2$ transitions 
to the ground state and by relatively strong $E1$ and $M1$ transitions 
to the isoscalar $2^+_1$ state. 
We apply the proton--neutron version of the Interacting Boson Model 
including quadrupole and octupole bosons ($sdf$--IBM-2).  
Two $F$-spin symmetric dynamical symmetry limits of the model, 
namely the vibrational and the $\gamma$-unstable ones, are considered.
We derive analytical formulae for excitation energies as well as 
$B(E1), B(M1), B(E2)$, and $B(E3)$ 
values for a number of transitions between low-lying states. 
\end{abstract}

\section{Introduction}

The phonon concept is a useful and simple concept in nuclear structure 
physics \cite{BM}. 
The low-lying collective $J^\pi = 2^+$ and $3^-$ excitations in 
near-spherical nuclei can be considered as quadrupole and 
octupole vibrations, which represent the most important vibrational 
degrees of freedom at low energies. 
The bosonic phonon concept suggests the occurrence of multiphonon 
states, which can decay collectively to states with one phonon 
less by the annihilation of one phonon. 
In a harmonic picture $n$-phonon states have excitation energies of 
$n$ times the one-phonon energy. 
Two of the appealing aspects of the study of multiphonon excitations 
high above the yrast-line is the investigation of their anharmonicities 
reflecting the influence of the Pauli principle and higher order 
residual interactions as well as the interaction of multiphonon modes 
with single particle degrees of freedom. 
Certainly these interactions can cause possible fragmentation of the idealized 
multiphonon modes. 
This fragmentation or the eventual observation of almost unfragmented 
multiphonon states can tell how well the considered phonon is 
an eigenmode of the nuclear dynamical system. 
To judge the deviation from the idealized multiphonon picture in reality 
obviously needs the formulation of this idealized situation and the 
investigation in how far the collective picture can account for the data.

Multiphonon and two-phonon states are at 
present a very actively investigated topic in nuclear structure physics. 
A two-quadrupole-phonon ($2^+ \otimes 2^+$) triplet with states 
having spin and parity quantum numbers $J^\pi = 0^+$, $2^+$, $4^+$ 
is usually known in near spherical nuclei. 
The absence of $J^\pi = 1^+, 3^+$ members of this two-phonon 
multiplet points at the bosonic character of the identical 
quadrupole phonons involved. 
Multi-quadrupole-phonon states (with $n>2$) have also been identified, 
see e.g. \cite{Appr87,Kern,CaJo92}. 
The investigation and identification of double giant dipole 
resonances (2GDR), ($3^-\otimes 3^-$) double-octupole states 
and ($2^+_\gamma \otimes 2^+_\gamma$) double-gamma-vibrations 
in deformed nuclei are actively debated in the literature, 
see e.g. Refs. \cite{ScBr93,AuEm98,VoCh99,KlSt82,YeGa96,BoJo91,GaKa97}.

The mixed-multipolarity ($2^+ \otimes 3^-$) quadrupole-octupole 
coupled quintuplet \cite{Lip66,Vog71} with spin 
and parity quantum numbers $J^\pi = 1^-$ -- $5^-$ is a very 
interesting example of the coupling of two different phonons. 
Members of the ($2^+ \otimes 3^-$) two-phonon quintuplet should 
decay by definition by collective $E2$ and $E3$ transitions to 
the $3^-_1$ octupole phonon state and to the $2^+_1$ 
quadrupole phonon state, respectively. 
The ($2^+ \otimes 3^-$) two-phonon states are expected to occur close 
to the sum energy $E_{2^+\otimes 3^-} = E(2^+_1) + E(3^-_1)$. 
There are only a few examples, e.g. $^{144}$Sm~\cite{Gat90}, 
$^{144}$Nd~\cite{Rob94,Hicks98}, $^{142}$Ce~\cite{Van95} and 
$^{112}$Cd~\cite{Gar99}, where besides the two-phonon $1^-$ state 
other multiplet members have been identified experimentally. 
The reason might be that all the other states from the $2^+ \otimes 3^-$ 
quintuplet, except the $1^-$, are strongly mixed with non-collective 
excitations, losing their pure quadrupole--octupole nature. 
This is supported by shell-model calculations 
performed recently for $N = 82$ nuclei~\cite{Ess99}. 
Thus, the most complete data, which can be used as a testing ground 
for the concept of quadrupole-octupole coupled two-phonon excitations, 
are the transition strengths from the mostly unfragmented $1^-$ member of 
the quintuplet.

The $J^\pi = 1^-$ two-phonon state has been very well investigated in 
magic and near spherical nuclei. 
In many nuclei, it has been identified close to the sum energy 
$E_{2^+\otimes 3^-}$, indicating a quite harmonic phonon coupling 
\cite{Metz78,Herz95,Brys99}. 
According to its definition, the two-phonon $1^-$ state should decay 
by a collective one-phonon $E2$ transition to the $3^-$ 
octupole phonon state. 
This has been confirmed experimentally \cite{Wil96,Wil98}. 
The collective $1^-_{2^+\otimes 3^-} \to 2^+_1$ $E3$ transition has 
not yet been measured due to the possibly dominating $E1$ and $M2$ 
contaminations in this transition. 
A particularly interesting feature of the two-phonon $1^-$ state is 
the existence of a relatively strong $E1$ transition to the ground 
state, which can be sensitively measured in photon scattering 
experiments \cite{KnPi96}. 
The strength of this two-phonon annihilating $E1$ transition 
approximately equals the strength of the $3^-_1 \to 2^+_1$ 
two-phonon changing $E1$ transition \cite{NPE199} which strongly 
supports quadrupole-octupole-collective origin of these $E1$ transitions.

All the low-energy one-phonon and two-phonon states mentioned above 
involve isoscalar phonons. 
Recently the isovector quadrupole excitation in the valence shell, 
the $2^+_{\rm ms}$ state, has been identified \cite{Ham84} in some 
nuclei from the measurement of absolute transition strengths, 
see, e.g., \cite{Hicks98,Van95,Ver88,FaBe92,GaLe96,WiGe97,Ba136,Mo94}. 
Isovector excitations in the valence shell form a whole class 
of low-lying collective states called, more precisely, 
mixed-symmetry states \cite{ArOt77,TakaPhD,Iach84}. 
The fundamental $2^+_{\rm ms}$ state decays by a weakly collective $E2$ 
transition to the ground state and by a strong $M1$ transition 
to the isoscalar $2^+_1$ state, with an $M1$ matrix element of 
the order of one nuclear magneton \cite{Iach84,NPCame}. 
In an harmonic phonon coupling scheme one can expect also the existence 
of mixed-symmetry two-phonon multiplets, that involve at least one 
excitation of the mixed-symmetry quadrupole phonon. 
Two symmetric--mixed-symmetric ($2^+_1\otimes 2^+_{\rm ms}$) 
two-quadrupole phonon excitations with 
positive parity, namely a $J^\pi = 1^+$ state (the scissors mode), 
and a $J^\pi = 3^+$ state have been 
identified already in the near spherical nucleus $^{94}$Mo \cite{Mo94,PiFr3+}. 
Similarly one can expect the existence of a ($2^+_{\rm ms} \otimes 3^-$) 
mixed-symmetry quadrupole-octupole coupled multiplet 
with negative parity. 
In a naive phonon coupling picture members of this quintuplet should 
show a very complex but, nevertheless,  simple 
to understand decay pattern, which can be predicted from the 
decay properties of the one-phonon excitations 
(weakly-collective $E2$ decay and strong $M1$ decay of the $2^+_{\rm ms}$ 
state, and collective $E3$ and relatively strong $E1$ decay of 
the $3^-_1$ octupole vibration). 
For instance, the $J^\pi = 1^-_{2^+_{\rm ms} \otimes 3^-}$ state should 
decay by relatively strong $E1$, strong $M1$, weakly-collective $E2$, 
and collective $E3$ transitions to the ground and to the $1^+$ scissors state, 
to the ordinary, isoscalar ($2^+ \otimes 3^-$) multiplet, and to the 
one-phonon $3^-$ and $2^+_{\rm ms}$ states, respectively (see Fig.~1).

Recently, $\gamma$ transitions between the $3^-$ octupole phonon 
state and the $2^+_{\rm ms}$ state have been observed experimentally 
\cite{Van95,Franpc} in near spherical nuclei. 
For an understanding of these new observations it is necessary to 
investigate the interrelation of the octupole phonon state and 
the mixed-symmetry $2^+_{\rm ms}$ state theoretically. 
That is one aim of the present article. 
Another purpose of this article is the proposal of possible 
two-phonon states generated by the coupling of the octupole 
degree of freedom with the proton-neutron degree of freedom 
and the quantitative investigation of the properties of such 
constructions.

A convenient and sometimes successful approach to new and complicated 
physical phenomena is the investigation of symmetries and the 
algebraic solution of symmetric Hamiltonians. 
In low-energy nuclear structure physics this task can be achieved 
by application of the algebraic interacting boson model 
\cite{ArIa75,IaAr87}. 
Since we deal simultaneously with the isoscalar octupole vibration 
and the isovector quadrupole vibration in the valence shell of near 
spherical nuclei, we need a model which incorporates them both.
We apply for the first time the $sdf$-IBM-2, 
which considers $l^\pi = 0^+$ monopole ($s$) bosons, 
$l^\pi = 2^+$ quadrupole ($d$) bosons, $l^\pi = 3^-$ octupole 
($f$) bosons, and the proton-neutron ($\pi$ - $\nu$) degree of 
freedom \cite{ArOt77}.

In this paper we develop the formalism of the $sdf$-IBM-2 in two particular
dynamical symmetry limits relevant for the description of vibrational and 
$\gamma $-unstable nuclei. 
In section \ref{sec:qn} we discuss the group reduction chains 
of the $sdf$-IBM-2 and the quantum numbers, which are necessary to classify 
the eigenstates of the dynamically symmetric Hamiltonians. 
We derive analytical formulae for $\gamma$ transition rates. 
The $E1$ transitions are calculated with a two-body operator. 
Quadrupole-octupole coupled two-body operators have turned out 
to be essential for the description of $E1$ transitions generated 
by quadrupole-octupole collectivity \cite{Han88,Barf89,NPE199}.

\section{$sdf$-IBM-2}
\label{sec:qn}

The building blocks of the $sdf$-IBM-2 
are $2 \times 13$ creation and $2 \times 13$ annihilation operators
\begin{equation}
\label{boson}
b^{+}_{\rho l m},  b_{\rho l m}
\end{equation} 
where $\rho =\pi , \nu $, $l=0,2,3$, $m=0, \pm 1, \ldots, \pm l$,
which by definition satisfy standard boson commutation relations:
\begin{equation}
\label{commutator}
\left[ b_{\rho l m}, b^{+}_{\rho' l' m'} \right] = 
\delta_{\rho \rho'} \delta_{l l'} \delta_{m m'} \; . 
\end{equation}     
The total number of bosons is conserved for a given nucleus, i.e.
\begin{equation}
\label{N}
N=N_{\pi} + N_{\nu }=n_{s \pi }+n_{d \pi }+n_{f \pi }
+n_{s \nu }+n_{d \nu }+n_{f \nu } \; .
\end{equation} 
The integers $n_{l \rho }$ in Eq. (\ref{N}) are the eigenvalues of the 
boson number operators 
$\hat{n}_{l \rho }=\left(b^+_{\rho l }\cdot \tilde b_{\rho l} \right)$,
where the dot denotes the scalar product  and  
$\tilde b_{\rho l m}= (-1)^{l-m} b_{\rho l, -m}$.
\par
The $2\times 13^2 = 338$ bilinear combinations of proton and neutron boson
operators of the type
\begin{equation}
\label{bilinear}
b^{+}_{\rho l m} b_{\rho l' m'}
\end{equation}     
generate the Lie algebra U$_{\pi }$(13) $\otimes $ U$_{\nu }$(13).

\subsection{Group reductions, quantum numbers and wave functions}

The U$_{\pi }$(13) $\otimes $ U$_{\nu }$(13) symmetry algebra has a rich
substructure. In the present study we are interested in the $F$-spin symmetric
dynamical symmetry limits of the model, which can be characterized by the
reduction chains
\begin{equation}
\label{DS}
\begin{array}{ccccccccccccccc}
{\mbox{U}_{\pi }(13)}&{\otimes }&{\mbox{U}_{\nu }(13)}&{\supset }&
{\mbox{U}_{\pi \nu }(13)}&{\supset }&
{\mbox{U}_{\pi \nu }(6)}&{\otimes }&{\mbox{U}_{\pi \nu }(7)} &
{ \supset \cdots \supset }&{\mbox{SO}^d_{\pi \nu }(3)}&
{\otimes }&{\mbox{SO}^f_{\pi \nu }(3)}&{\supset }&{\mbox{SO}_{\pi \nu }(3)} \\ 
\downarrow & & \downarrow & & \downarrow & & \downarrow & & \downarrow & & 
\downarrow & & \downarrow & & \downarrow \\[0pt]     
{[N_{\pi }]} & & {[N_{\nu }]} & & {[N_1,N_2]} & &
{[n_1,n_2]} & & {[m_1,m_2]} & {\ldots} & L_d & & L_f & & L \\     
\end{array}
\end{equation}              
where the dots denote alternative group reductions discussed later. 
Below each group we indicate the quantum numbers, associated with it.
In total, the complete classification of the basis states  (\ref{DS})
requires 26 quantum numbers corresponding to the $2\times 13$ 
occupation numbers $n_{\rho l m}$ in the $m$-scheme. 
The $F$-spin quantum number~\cite{ArOt77,IaAr87} relates in terms of 
Young tableaux to the one- or
two-rowed U$_{\pi \nu }$(13) representation as 
$$
F=\frac{N_1-N_2}{2} \; .
$$
The maximum value of the $F$-spin quantum number 
$$
F_{\rm max}=\frac{N}{2} 
$$
corresponds to the totally symmetric representations of the 
U$_{\pi \nu }$(13) group. 
States with smaller $F$-spin quantum numbers $F<F_{\rm max}$ 
are called mixed-symmetry states and correspond to 
non-symmetric representations of the U$_{\pi \nu }$(13) group. 
Since the $sdf$-IBM-2 contains an octupole degree of freedom,
it enlarges the diversity of mixed-symmetry states compared to the standard 
$sd$-IBM-2.
This fact can be recognized from the branching rules of
U$_{\pi \nu }$(13) $\supset $ U$_{\pi \nu }$(6) $\otimes $ U$_{\pi \nu }$(7)
(see Appendix A).

The simplest two-rowed representation of U$_{\pi \nu }$(13) reduces
to the U$_{\pi \nu }$(6) $\supset $ U$_{\pi \nu }$(7) representations
in the following way: 
\begin{mathletters} 
\begin{eqnarray}
\nonumber
{\rm U}_{\pi \nu }(13) & \supset  &
{\rm U}_{\pi \nu }(6)\otimes {\rm U}_{\pi \nu }(7) \\[2pt]
\eqnum{6.1}\label{6.1} 
[N-1,1]& = & [n-1,1]\otimes [m] \\    
\eqnum{6.2}\label{6.2} 
       &   & [n]\otimes [m-1,1] \\
\eqnum{6.3}\label{6.3} 
       &   & [n]\otimes [m] 
\end{eqnarray}  
\end{mathletters} 
where, $n=n_1 + n_2$ is the total number of $s$ and $d$ bosons,
$m=m_1 + m_2$ is the number of $f$ bosons, giving thus $N=n+m$.

As follows from Eq. (6), three types of the mixed-symmetry states arise.
\begin{enumerate}
\item  The $[n-1,1] \otimes [m]$ decomposition corresponds to the usual
mixed-symmetry states in the $sd$-space,
coupled to $m$ octupole-bosons with a wave function, 
which is symmetric in the $f$-sector. 
Examples are the fundamental $2^+_{\rm ms}$ state and the $1^+_{sc}$ 
``scissors'' state 
with no $f$-boson at all, which belong to the 
${\rm U}_{\pi \nu }(6)\otimes {\rm U}_{\pi \nu }(7)$ representation 
$[n-1,1] \otimes [0]$, 
or the new negative-parity mixed-symmetry two-phonon states which 
we denote as $(2^+_{\rm ms} \otimes 3^-_1)$ because they are generated by the 
coupling of the lowest $2^+_{\rm ms}$ state in the $sd$-sector and one 
symmetric $f$-boson, which belong to the 
${\rm U}_{\pi \nu }(6)\otimes {\rm U}_{\pi \nu }(7)$ representation 
$[n-1,1] \otimes [1]$. 
Note that in the presence of one $f$-boson due to the 
boson number conservation only $n = N-1$ bosons remain in the 
$sd$-sector, which can belong to the symmetric 
${\rm U}_{\pi \nu }(6)$ representation $[n] = [N-1]$ or 
to the mixed-symmetry representations, i.e., 
to the lowest $[n-1,1] = [N-2,1]$. 
Therefore, the representation $[n-1,1] \otimes [1]$ can be considered 
as $(2^+_{\rm ms} \otimes 3^-_1)$ two-phonon states. 
It turns out from the application of the branching rules given 
in Appendix A that this $(2^+_{\rm ms} \otimes 3^-_1)$ two-phonon coupling 
will generate mixed-symmetry states with $F$-spin quantum numbers 
$F= F_{\rm max}-1$ and $F= F_{\rm max}-2$, 
because the representation $[N-2,1] \otimes [1]$ is also present 
in the U$_{\pi\nu}$(13) representation $[N-2,2]$, which has a total 
$F$-spin quantum number $F=F_{\rm max}-2$. 
In this article, we restrict ourselves to mixed-symmetry states with 
$F= F_{\rm max}-1$.

\item  The $[n]\otimes [m-1,1]$ reduction describes the mixed-symmetry states
which appear due to mixed-symmetry in the $f$-sector. 
We will not consider such kind of states in the present paper.

\item  Finally, $[n]\otimes [m]$ states are symmetric separately 
in the $sd$- and in the $f$-sectors, however, they are coupled 
in a non-symmetric way within the full $sdf$-space. 
The simplest example for a mixed-symmetry state if this type is the 
$3^-_{\rm ms}$ state, which consists of $m=1$ $f$-boson and 
$n=N-1$ $s$-bosons. 
This $3^-_{\rm ms}$ state is the mixed-symmetry analogue state of the 
symmetric octupole phonon, the $3^-_1$  state. 
This can be clearly seen from the explicit wave functions given 
in Table \ref{tab:WvfctsU5}. 
Higher excited mixed-symmetry states of this type can be obtained by 
replacing in a symmetric way $s$-bosons with $d$-bosons in 
the $3^-_{\rm ms}$ wave function along with proper normalization. 
We denote the lowest energy example of such mixed-symmetry states 
schematically as $(2^+ \otimes 3^-_{\rm ms})$.
\end{enumerate}

The operator which controls the excitation energy 
of mixed-symmetry states with $F$-spin quantum numbers 
$F < F_{\rm max}$ is the Majorana operator.
By definition, this is the most general operator which 
annihilates any totally symmetric state.
Within the  U$_{\pi \nu }$(13) $\supset $ 
U$_{\pi \nu }$(6) $\otimes $ U$_{\pi \nu }$(7),
there are three important types of Majorana operators.

The well known Majorana operator $\hat{M}_6$~\cite{ArOt77} is 
associated with the U(6) group generated by the $s$- and $d$-bosons only. 
It has the form 
\begin{equation} 
\label{eq:Majo6}
\hat{M}_6=[s^+_\nu \times d^+_\pi -s^+_\pi \times d^+_\nu ]^{(2)} 
 \cdot[s_\nu \times \tilde d_\pi -s_\pi \times \tilde d_\nu ]^{(2)} 
  - 2 \sum_{k=1,3} [d^+_\nu \times d^+_\pi ]^{(k)} \cdot 
       [\tilde d_\nu \times \tilde d_\pi ]^{(k)} \; ,
\end{equation}
which influences the states whose wave functions are non-symmetric 
in the $sd$-sector of the model space.
%\begin{equation} 
%\label{state1}
%2^+_{\rm ms} & \sim & (d^+_\pi s_\pi - d^+_\nu s_\nu ) |{\rm g.s.} \rangle \\
%\end{eqnarray}
In Eq.~(\ref{eq:Majo6}) and below, $\times $ denotes 
the standard tensor product of two irreducible tensorial operators. 
It relates to the ${\rm U}_{\pi \nu }(6)$ quadratic Casimir operator as
\begin{equation}
\label{M6}
C_2[\mbox{U}_{\pi \nu }(6)]=n(n+5)-2 \hat{M}_6 \; . 
\end{equation}

\noindent
The Majorana operator $\hat{M}_7$ is 
associated with the U(7) group generated by the $f$-bosons only. 
It has the form 
\begin{equation} 
\label{eq:Majo7}
\hat{M}_7= -2 \sum_{k=1,3,5} [f^+_\nu \times f^+_\pi ]^{(k)} \cdot 
        [\tilde f_\nu \times \tilde f_\pi ]^{(k)} \
\end{equation}
which is responsible for pushing up the mixed-symmetry states of 
the second type, i.e non-symmetric in the $f$-sector.
It relates to the ${\rm U}_{\pi \nu }(7)$ quadratic Casimir invariant as
\begin{equation}
\label{M7}
C_2[\mbox{U}_{\pi \nu }(7)]=m(m+6)-2 \hat{M}_7 \; . 
\end{equation}

\noindent
Finally, the Majorana operator which is associated with the full 
group U(13), reads
\begin{eqnarray} 
\label{eq:Majo13}
\hat{M}_{13} &=& \hat{M}_6 + \hat{M}_7 + 
[s^+_\nu \times f^+_\pi -s^+_\pi \times f^+_\nu ]^{(3)} 
 \cdot[s_\nu \times \tilde f_\pi -s_\pi \times \tilde f_\nu ]^{(3)} \\
& &+\sum_{k=1,2,3,4,5} (-1)^{k+1} 
[d^+_\nu \times f^+_\pi -d^+_\pi \times f^+_\nu ]^{(k)} \cdot 
[\tilde d_\nu \times \tilde f_\pi -\tilde d_\pi \times \tilde f_\nu ]^{(k)} 
\; ,\\ 
   & = & \left[F_{\rm max}\left(F_{\rm max}+1\right) - \hat{F}^2 \right] \; .
\end{eqnarray}    
This operator acts on all three types of mixed-symmetry states.
$\hat{M}_{13}$ relates to the ${\rm U}_{\pi \nu }(13)$
quadratic Casimir operator as
\begin{equation}
\label{M13}
C_2[\mbox{U}_{\pi \nu }(13)]=N(N+12)-2 \hat{M}_{13} \; .
\end{equation}

The most general Hamiltonian constructed from linear and quadratic 
Casimir operators of the group U$_{\pi \nu }$(6) $\otimes $ U$_{\pi \nu }$(7) 
reads 
\begin{equation}
\label{HU6U7}
\begin{array}{ll}
H   & = H_0+ \lambda \hat{M}_{13}+ \lambda' \hat{M}_6 + \lambda ''\hat{M}_7    
      +\epsilon_dC_1[\mbox{U}_{\pi \nu }(5)] 
      +\alpha C_2[\mbox{U}_{\pi \nu }(5)] 
      +\kappa^\prime C_2[\mbox{SU}_{\pi \nu }(3)]\\ 

    & +\eta C_2[\mbox{SO}_{\pi \nu }(6)]
      +\beta C_2[\mbox{SO}_{\pi \nu }(5)]
      +\gamma_dC_2[\mbox{SO}^d_{\pi \nu }(3)] 
      +\epsilon_fC_1[\mbox{U}_{\pi \nu }(7)] \\

    & +\kappa C_2[\mbox{SO}_{\pi \nu }(7)] +\xi C_2[\mbox{G}_{2 \pi \nu }] 
      +\gamma_fC_2[\mbox{SO}^f_{\pi \nu }(3)]
      +\gamma C_2[\mbox{SO}_{\pi \nu }(3)]\;.  
\end{array}
\end{equation}  
The definitions of the Casimir invariants of Lie groups used here
are given in Appendix B.   
The structure of the Majorana operators has interesting consequences. 
The $sdf$-IBM-2 reduces to the simpler version $sdf$-IBM-1, if 
the strength constant $\lambda$ of the Majorana operator $\hat{M}_{13}$ 
is set to infinity. 
Using a finite value of $\lambda$, but $\lambda '' = \infty$ will remove 
only those states which have mixed-symmetry in the $f$-sector alone. 
We will use this choice throughout the paper. 
For the ordinary mixed-symmetry states in the $sd$-sector, which 
are known already from the $sd$-IBM-2, the sum $\lambda + \lambda^\prime$ 
plays the role of the ordinary Majorana parameter in the $sd$-IBM-2. 
The excitation energies of mixed-symmetry states, 
which contain one $f$-boson are in addition 
also functions of the quantity $\lambda - \lambda^\prime$. 
This parameter controls the excitation energy of mixed-symmetry states 
with negative parity.

Since we are going to study vibrational or $\gamma $-unstable nuclei, we are
interested in two well-known dynamical symmetry limits of the positive parity
$sd$-subchain \cite{IaAr87}, namely, 
\begin{equation}
\label{DS1}
\mbox{U}_{\pi \nu }(6) \supset 
\left\{
\begin{array}{c}
\mbox{U}_{\pi \nu }(5) \\
\mbox{SO}_{\pi \nu }(6) \\
\end{array} 
 \right\}  
\supset \mbox{SO}_{\pi \nu }(5)\supset \mbox{SO}_{\pi \nu }(3) 
\end{equation}    
which we consider in the following subsections.

The negative parity subchain has a unique reduction~\cite{Racah,Roh78}:
\begin{equation}
\label{DS2}
\begin{array}{ccccccc}
\mbox{U}_{\pi \nu }(7) & \supset & \mbox{SO}_{\pi \nu }(7) & \supset &
\mbox{G}_{2 \pi \nu } & \supset & \mbox{SO}_{\pi \nu }(3) \\
\downarrow & & \downarrow & & \downarrow & & \downarrow  \\[0pt]     
[m_1,m_2] & & (\omega_1,\omega_2) & & (u_1,u_2) & \{\beta_j\} & L_f  \\  
\end{array}   
\end{equation} 
where the quantum numbers $\{\beta_j\}$ ($j=1,2,3,4$) are 
the so-called missing labels, which are necessary
to classify completely the G$_2 \supset $ SO(3) reduction.
Unlike the case of $sdf$-IBM-1 where only the totally symmetric 
representations
are important~\cite{ArIa75,Roh78}, we also deal here with two-rowed 
representations of the algebras in (\ref{DS}) and (\ref{DS2}). 
This requires some modifications.
For example, we have to take into account the exceptional group G$_2$.  
For the reduction of not fully symmetric representations of SO(7) the group 
G$_2$ helps to resolve the labelling problem. 
The missing label, which is in general necessary to uniquely 
define the SO(7) $\supset$ G$_2$ reduction, is redundant in our 
case of only two-rowed representations.

\subsubsection{The U$_{\pi \nu }$(1)$\otimes $U$_{\pi \nu }$(5)$\otimes 
$U$_{\pi \nu }$(7) dynamical symmetry limit}

In this dynamical symmetry limit the reduction chain of the positive parity 
U$_{\pi \nu }$(6) subalgebra is~\cite{IaAr87}
\begin{equation}
\label{DSU5}
\begin{array}{ccccccc}
\mbox{U}_{\pi \nu }(6) & \supset & \mbox{U}_{\pi \nu }(5) & \supset & 
\mbox{SO}_{\pi \nu }(5) & \supset & \mbox{SO}^d_{\pi \nu }(3) \\ 
\downarrow & & \downarrow  & & \downarrow & & \downarrow \\[0pt]     
[n_1,n_2] & & \{n_{d_1},n_{d_2}\} & & [\tau_1,\tau_2] & \{\alpha_i\} & L_d \\     
\end{array}
\end{equation}              
where $\alpha_i $ ($i=1,2$) are missing labels, necessary to 
completely classify the SO(5) $\supset $SO(3) reduction. 
We denote here $n=n_1 + n_2$ the total number of $s$ and $d$ bosons,
while the number of $d$ bosons only is denoted as $n_d=n_{d_1}+n_{d_2}$.
Thus, the total wave function can be written as
\begin{equation}
\label{wfU5}
{|[N_{\pi }] {\otimes }[N_{\nu }];[N_1,N_2][n_1,n_2]\{n_{d_1},n_{d_2}\}
(\tau_1,\tau_2)\{\alpha_i \} L_d;[m_1,m_2](\omega_1,\omega_2)(u_1,u_2)
\{\beta_j\} L_f;LM \rangle }                
\end{equation}
The quantum numbers of the lowest states 
are given in Table I.

In the U$_{\pi \nu }$(5)$\otimes $U$_{\pi \nu }$(7) dynamical symmetry limit 
it is rather simple to construct explicitly the wave functions.
We shall do this in order to make clear the variety of mixed-symmetry states 
appearing in the model.
In Table I we present the wave functions for some states of interest.
We discussed above already the one-$f$-boson octupole states. 
Of particular interest are here also the lowest $1^-$ states. 
From the coupling of one $d$-boson quadrupole phonon and
one $f$-boson octupole phonon four $1^-$ states emerge. 
This is evident in the $m$-scheme in $F$-space because the four 
basis wave functions $|d_\pi f_\pi\rangle$, $|d_\nu f_\pi\rangle$, 
$|d_\pi f_\nu\rangle$, and $|d_\nu f_\nu\rangle$ can be formed. 
These basis states can be combined to states with U$_{\pi \nu }$(13) 
symmetry and good $F$-spin. 
One obtains one symmetric $1^-$ state
corresponding to the totally symmetric $[N]$ irrep of U$_{\pi \nu }$(13) 
with $F$-spin quantum number $F=F_{\rm max}$, 
and three mixed-symmetry states: two of them with $F$-spin quantum 
numbers $F=F_{\rm max}-1$ and one with $F=F_{\rm max}-2$. 
One $F=F_{\rm max}-1$ state corresponds to the 
$[N-1,1] \supset [N-2,1]\otimes [1]$ reduction and the other 
to the $[N-1,1] \supset [N-1]\otimes [1]$ reduction in the 
U$_{\pi \nu }$(13)$\supset $ U$_{\pi \nu }$(6)$\otimes $ U$_{\pi \nu }$(7),
chain. 
The last one arises from the $[N-2,2] \supset [N-2,1]\otimes [1]$ 
reduction. 
According to the discussion above we denote them as 
$1^- = (2^+\otimes 3^-)_{1^-}$, 
$1^-_{\rm ms} = (2^+_{\rm ms}\otimes 3^-)_{1^-}$, 
$1^{- \prime }_{\rm ms} = (2^+\otimes 3^-_{\rm ms})_{1^-}$ and 
$1^{- \prime \prime }_{\rm ms}$, respectively.

The  U$_{\pi \nu }$(5)$\otimes $U$_{\pi \nu }$(7) dynamical symmetry 
Hamiltonian expressed in terms of Casimir invariants of first 
and second order of the relevant groups 
reads
\begin{equation}
\label{HU5}
\begin{array}{ll}
H&=H_0+ \lambda \hat{M}_{13}+ \lambda' \hat{M}_6 + \lambda ''\hat{M}_7    
+\epsilon_dC_1[\mbox{U}_{\pi \nu }(5)] \\
&+\alpha C_2[\mbox{U}_{\pi \nu }(5)]   
+\beta C_2[\mbox{SO}_{\pi \nu }(5)]+\gamma_dC_2[\mbox{SO}^d_{\pi \nu }(3)] 
+\epsilon_fC_1[\mbox{U}_{\pi \nu }(7)] \\
&+\kappa C_2[\mbox{SO}_{\pi \nu }(7)] +\xi C_2[\mbox{G}_{2 \pi \nu }] 
+\gamma_fC_2[\mbox{SO}^f_{\pi \nu }(3)]
+\gamma C_2[\mbox{SO}_{\pi \nu }(3)]\;.  
\end{array}
\end{equation}

The eigenvalue problem for the dynamical symmetry Hamiltonian 
(\ref{HU5}) can be solved analytically. 
The solution of the full U(5) Hamiltonian reads 
\begin{equation}
\label{EU5}
\begin{array}{ll}
E&=E_0+\lambda\left[F_{\rm max}\left(F_{\rm max}+1\right)-F(F+1)\right]
+ \lambda' n_2 (n_1 + 1) + \lambda'' m_2 (m_1 + 1) \\
&+\epsilon_d(n_{d_1}+n_{d_2})+\alpha [n_{d_1}(n_{d_1}+4)+n_{d_2}(n_{d_2}+2)]  
+\beta [\tau_1(\tau_1+3)+\tau_2(\tau_2+1)]\\
&+\gamma_d L_d(L_d+1)+\epsilon_f(m_1+m_2)  
+\kappa [\omega_1(\omega_1+5)+\omega_2(\omega_2+3)] \\
&+\xi [u_1(u_1+5)+u_2(u_2+4)+u_1 u_2]+\gamma_fL_f(L_f+1)+ \gamma L(L+1) \; .   
\end{array}
\end{equation} 
To construct the spectrum one needs to know the branching rules for the groups
involved (see Appendix). 
An energy spectrum corresponding to (\ref{HU5}) is shown in
Fig.~2. The parameters of the Hamiltonian are specified 
in the figure caption.

\subsubsection{The SO$_{\pi \nu }$(6)$\otimes $U$_{\pi \nu }$(7) 
dynamical symmetry limit}

In this dynamical symmetry limit the reduction chain of the positive parity 
U$_{\pi \nu }$(6) subalgebra is
\begin{equation}
\label{DSSO6}
\begin{array}{ccccccc}
\mbox{U}_{\pi \nu }(6) & \supset & \mbox{SO}_{\pi \nu }(6) & \supset & 
\mbox{SO}_{\pi \nu }(5) & \supset & \mbox{SO}^d_{\pi \nu }(3) \\ 
\downarrow & & \downarrow  & & \downarrow & & \downarrow \\[0pt]     
[n_1,n_2] & & \langle \sigma_1,\sigma_2 \rangle & & [\tau_1,\tau_2] & 
\{\alpha_i\} & L_d \\     
\end{array}
\end{equation}              
and the total wave function is as follows:
\begin{equation}
\label{wfSO6}
{|[N_{\pi }] {\otimes }[N_{\nu }];[N_1,N_2] [n_1,n_2] 
\langle \sigma_1,\sigma_2 \rangle (\tau_1,\tau_2)\{\alpha_i\} L_d;
[m_1,m_2](\omega_1,\omega_2)(u_1,u_2)\{\beta_j\} L_f;LM \rangle  }       
\end{equation}   
The quantum numbers for the lowest states in this dynamical symmetry 
are presented in Table II.

The SO$_{\pi \nu }$(6)$\otimes $U$_{\pi \nu }$(7) 
dynamical symmetry Hamiltonian reads
\begin{equation}
\label{HSO6}
\begin{array}{ll}
H&=H_0+\lambda \hat{M}_{13}+ \lambda' \hat{M}_6 + \lambda ''\hat{M}_7   
+\zeta C_2[\mbox{SO}_{\pi \nu }(6)]\\  
&+\beta C_2[\mbox{SO}_{\pi \nu }(5)]+\gamma_dC_2[\mbox{SO}^d_{\pi \nu }(3)]
+\epsilon_fC_1[\mbox{U}_{\pi \nu }(7)] \\ 
&+\kappa C_2[\mbox{SO}_{\pi \nu }(7)]+\xi C_2[\mbox{G}_{2 \pi \nu }]
+\gamma_fC_2[\mbox{SO}^f_{\pi \nu }(3)]+\gamma C_2[\mbox{SO}_{\pi \nu }(3)]\;. 
\end{array}
\end{equation}     
and has the eigenvalues 
\begin{equation}
\label{ESO6}
\begin{array}{ll}
E&=E_0+\lambda\left[F_{\rm max}\left(F_{\rm max}+1\right)-F(F+1)\right]
+\lambda' n_2 (n_1 + 1) + \lambda'' m_2 (m_1 + 1) \\
&+\zeta [\sigma_1(\sigma_1+4)+\sigma_2(\sigma_2+2)]  
+\beta [\tau_1(\tau_1+3)+\tau_2(\tau_2+1)]
+\gamma_d L_d(L_d+1) \\
&+\epsilon_f(m_1+m_2)+\kappa [\omega_1(\omega_1+5)+\omega_2(\omega_2+3)] \\
& +\xi [u_1(u_1+5)+u_2(u_2+4)+u_1 u_2] 
+\gamma_fL_f(L_f+1)+ \gamma L(L+1) \; .   
\end{array}
\end{equation} 
The branching rules for the corresponding groups can be found in 
Ref.~\cite{PVI86} and in the Appendix below.
A typical energy spectrum corresponding to Eq. (\ref{ESO6}) is shown in
Fig.~3. The parameters are specified in the caption.

\subsection{Electromagnetic transitions}

In this subsection we discuss electromagnetic transitions between low-lying 
states in the $sdf$-IBM-2 with particular emphasis of those 
transitions which are new in the present model with respect 
to the standard $sd$-IBM-2 and $sdf$-IBM-1 species. 
For the study of the interrelation between the $F = F_{\rm max}$ 
octupole phonon and the $F= F_{\rm max} -1$ mixed-symmetry quadrupole phonon 
and possible two-phonon states generated by the coupling of them, 
we are interested in $E1, M1, E2$, and $E3$ transition strengths. 
The reduced transition probabilities 
\begin{equation}
\label{BEML}
B(\Pi \lambda; J_i \to J_f)=
\frac{1}{2J_i+1}|\langle J_f||T(\Pi \lambda )|| J_i \rangle |^2
\end{equation}  
are calculated from the matrix elements of the transition operators 
$T(\Pi \lambda )$, 
where $\Pi $ stands for the radiation character, either $E$ or $M$, 
and $\lambda$ denotes the multipolarity.

The electric quadrupole operator reads
\begin{equation}
\label{TE2}
T(E2)=e_{\pi } Q^{(2)}_{\pi } + e_{\nu } Q^{(2)}_{\nu } \; ,
\end{equation}      
where
\begin{equation}
\label{Q}
Q^{(2)}_{\rho }=[d^{+}_{\rho } \times s_{\rho } 
+ s^{+}_{\rho } \times \tilde d_{\rho }]^{(2)}
+\chi_{\rho }[d^{+}_{\rho } \times \tilde d_{\rho }]^{(2)} 
+\chi'_{\rho }[f^{+}_{\rho } \times \tilde f_{\rho }]^{(2)} \; . 
\end{equation}

The magnetic dipole operator has the form
\begin{equation}
\label{TM1}
T(M1)=\sqrt{\frac{3}{4 \pi }} \left(g_{\pi } L^d_{\pi } + g_{\nu } L^d_{\nu } 
+g'_{\pi } L^f_{\pi } + g'_{\nu } L^f_{\nu } \right) \; ,
\end{equation}      
where
\begin{equation}
\label{L}
L^d_{\rho }=\sqrt{10} [d^{+}_{\rho } \times \tilde d_{\rho }]^{(1)} \;, \quad
L^f_{\rho }=2\sqrt{7} [f^{+}_{\rho } \times \tilde f_{\rho }]^{(1)} \; .
\end{equation}

According to the Eqs. (\ref{TE2}--\ref{L}) the addition of the $f$ 
boson enlarges the structure of electromagnetic operators
of the standard $sd$-IBM-2. 
The standard $sd$-IBM-2, however, works  well for the description of 
$E2$ and $M1$ transitions. 
In order to reduce the number of model parameters in the $sdf$-IBM-2 
we, therefore, neglect below those parts of the $E2$ and $M1$ transition 
operators which involve the $f$-boson. 
Thus, we assume $\chi^\prime_\rho = 0$ and 
$g^\prime_\rho =0$. 
In contrast, the $f$-boson is essential for $E3$ and $E1$ transitions, which 
are new with respect to the $sd$-IBM-2.

Since there is at present no experimental evidence for the existence of 
states with mixed-symmetry in the $f$-boson sector, we shall consider 
here only $F$-scalar octupole transitions, 
i.e., we use the following symmetric octupole operator:
\begin{equation}
\label{TE3}
T(E3)=e_3 \left( O^{(3)}_{\pi } + O^{(3)}_{\nu } \right) \; ,
\end{equation}      
where $O^{(3)}_{\rho }$ stands for
\begin{equation}
\label{O}
O^{(3)}_{\rho }=[s^{+}_{\rho } \times \tilde f_{\rho }  
+ f^{+}_{\rho } \times s_{\rho }]^{(3)} 
+\chi [d^{+}_{\rho } \times \tilde f_{\rho }  
+ f^{+}_{\rho } \times \tilde d_{\rho }]^{(3)} \; . 
\end{equation}

The general one-body electric dipole operator has the form
\begin{equation}
\label{TE11body}
T(E1)=\alpha_{\pi } D^{(1)}_{\pi } + \alpha_{\nu } D^{(1)}_{\nu } \; ,
\end{equation}                               
where
\begin{equation}  
\label{D}
D^{(1)}_{\rho }=\left[ d^{+}_{\rho } \times \tilde f_{\rho }  
+ f^{+}_{\rho } \times \tilde d_{\rho } \right]^{(1)} \; . \\ 
\end{equation}

While the one-body $M1$, $E2$, and $E3$ transition operators work 
satisfactorily, the one-body $E1$-transition operator fails to 
provide an adequate description of low-lying dipole transition strengths 
in vibrational nuclei~\cite{Han88}. 
This applies for near-spherical nuclei also in the case when a $p$-boson 
is considered as it will be discussed below. 
The operator (\ref{TE11body}) has very strict selection rules, 
conserving the total number of $d$- and $f$-bosons $\Delta(n_d + n_f)=0$. 
This contradicts the experimental findings where the 
quadrupole-octupole collective $E1$ transitions, including the 
ground state decay of the quadrupole-octupole two-phonon $1^-$ state, 
are even found to be enhanced, 
see e.g. Refs. \cite{Metz78,Herz95,Brys99,KnPi96}.

Two approaches exist to avoid this problem.
Following the first, the $sdf$-IBM is enlarged with a $p$ boson 
($l^{\pi }=1^-$), thus leading to the so-called $spdf$-IBM 
\cite{MaVe85,Eng87,OtSu88,Kus88}. 
The inclusion of the $p$ boson in the IBM corresponds to the 
consideration of a collective dipole vibration. 
There exists no unambiguous interpretation of the $p$-boson origin.
Some authors point out the origin of the $p$-boson from the collective
low-lying valence shell excitation after removing the center-of-mass motion
\cite{Eng87,Kus88}. 
Others have associated the $p$-boson with the GDR \cite{MaVe85,OtSu88}. 
The $p$ boson induced one-body electric dipole operator works 
succesfully for the description of dipole transition strengths in 
deformed nuclei \cite{MaVe85,Eng87,OtSu88,Kus88,SuOt96,Long98}.
For near-spherical nuclei, where the present article 
is focussing on, the inclusion of the $p$ boson does not lift 
all discrepancies~\cite{Gar99}.

The lowest $1^-$ state is the only strong low-lying $E1$ excitation 
in many near-spherical nuclei \cite{KnPi96}. 
This fact strongly suggests that this state is identified with the 
collective negative-parity dipole mode in the case of considering 
a low-energy $p$ boson. 
This view, however, renders the well-established agreement 
of the $1^-$ excitation energy with the quadrupole-octupole two-phonon 
sum energy a sheer coincidence. 
Also the enhanced $E1$ strengths from other members of the 
quadrupole-octupole coupled quintuplet with spin quantum 
numbers $J\neq 1$ are not at all simple to understand.

On the other hand some other experimental facts indicate different 
origin of strong $E1$ transitions in near-spherical nuclei than sheer 
mixing with the GDR. 
For example, studies of $E1$-strength distribution in 
$^{140}$Ce~\cite{HeBr97} report that the lowest $1^-$ 
state exhibits an enhanced dipole transition to the ground 
state, while transitions from a number of higher lying 
$1^-$ states are considerably weaker although they lie 
closer to the GDR and although they are calculated to have 
predominantly a simple 1p1h-structure, which would even favor 
a mixing with the GDR. 
Thus, the observed concentration of $E1$-strength in the lowest
$1^-$ states should not be simply attributed to admixtures of the 
GDR into the states wave function if not at the same time the 
smaller $E1$ strengths of higher lying $1^-$ states are explained, too. 
It is even more striking that a model which can treat all these $E1$ 
excitations on the same footing, the microscopic Quasiparticle Phonon Model, 
supports the interpretation of enhanced low-lying $E1$ transitions due to 
quadrupole-octupole collectivity \cite{PonoE1}.

Therefore, another approach is preferable. 
The multiphonon structure of excitations in vibrational nuclei as well 
as an octupole or quadrupole-octupole coupled origin of negative-parity 
states, naturally suggests an $E1$-transition operator which 
contains a quadrupole-octupole coupled two-body part. 
Such a picture is also supported by the empirical correlation 
between $E1$-transition strengths of the  $1^-_1 \to 0^+_1$ and 
$3^-_1 \to 2^+_1$ transitions found recently for heavy near-spherical 
nuclei~\cite{NPE199} and by the fore-mentioned microscopic calculations 
\cite{PonoE1}. 
Furthermore, it avoids the problems in connection with 
the introduction of a $p$ boson discussed above. 
We choose, therefore, to work with a quadrupole-octupole coupled 
two-body part in the $E1$ operator.

The two-body dipole operator has already been introduced within the simple 
quadrupole and octupole phonon model~\cite{NPE199} and earlier within 
the $sdf$-IBM-1 \cite{Barf89,Bren92}. 
This ansatz is in good agreement with the available data. 
The $E1$ transition operator considered here has the form
\begin{equation}
\label{TE1}
\begin{array}{c}
\displaystyle T(E1)=\alpha_{\pi } D^{(1)}_{\pi } + 
\alpha_{\nu } D^{(1)}_{\nu } + \\[3pt]
\displaystyle \frac{\beta }{2N}
\left[ \left(Q^{(2)}_{\pi }{+}\eta Q^{(2)}_{\nu }\right) 
{\times }\left(O^{(3)}_{\pi }{+}O^{(3)}_{\nu }\right){+} 
\left(O^{(3)}_{\pi }{+}O^{(3)}_{\nu }\right){\times }
\left(Q^{(2)}_{\pi }{+}\eta Q^{(2)}_{\nu }\right) \right]^{(1)} \; ,
\end{array}
\end{equation}                               
where $D^{(1)}_\rho$, $Q^{(2)}_\rho$, and $O^{(3)}_\rho$ are given by the 
Eqs. (\ref{D}), (\ref{Q}) with $\chi'_{\rho }=0$, and (\ref{O}), respectively,
and $\eta =e_{\nu }/e_{\pi }$. 
We stress that we introduce only one free parameter more ($\beta$) 
in the $E1$ operator with respect to the one-body operator from 
Eq. (\ref{TE11body}). 
Below we analyze $E1$-transitions within
two particular dynamical symmetry limits.

\subsubsection{The U$_{\pi \nu }$(1)$\otimes $U$_{\pi \nu }$(5)$\otimes 
$U$_{\pi \nu }$(7) dynamical symmetry limit}

Using the wave functions (\ref{wfU5}) we can obtain the matrix elements of the 
transition operators (\ref{TE2}), (\ref{TM1}), (\ref{TE3}) and (\ref{TE1}).
The technique for the calculation of such matrix elements 
is described in detail in Ref. \cite{PVI86}. 
In this particular dynamical symmetry limit, it is more convenient 
to use the explicit expressions of the wave functions 
constructed from boson operators  
following the method described in \cite{Sco80} (see, e.g., Ref.~\cite{PVI86}).
The $E2$ and $M1$ transition strengths between those states, whose 
wave functions do not involve any $f$ bosons, are exactly the same 
as in the case of the $sd$-IBM-2 and can be found in \cite{PVI86}.  
The advantage of the $sdf$-IBM-2 is that it provides the description 
of $E2$ and $M1$ transitions between negative parity states and 
of $E1$ and $E3$ transitions between the states of different parity 
and different $F$-spin. 
Here we are particularly interested in the electric dipole transitions 
which connect 
the octupole $3_1^-$ state with the symmetric $2^+_1$ state and 
with the mixed-symmetry $2_{\rm ms}^+$ state, 
as well as in those which connect the $(2^+ \otimes 3^-)$ two-phonon 
negative parity multiplet with the ground state band.
We also consider here the new mixed-symmetry quadrupole-octupole coupled 
dipole states $1^-_{\rm ms}=(2^+_{\rm ms}\otimes 3^-)_{1^-}$ and
$1^{-\prime }_{\rm ms}=(2^+ \otimes 3^-_{\rm ms})_{1^-}$ and 
their electromagnetic properties (including $M1$ strengths) as 
examples for mixed-symmetry states with negative parity. 
Analytical expressions of reduced transition probabilities 
are summarized in Table III.

It is worthwhile to look closely to the analytical expressions 
in Table III in order recognize the fingerprints of the simple 
phonon picture the consideration of which has guided us to 
the present model. 
The $E1$ transition strengths from the $3^-_1$ state to the symmetric 
and mixed-symmetry one-phonon quadrupole excitations, the $2^+_1$ state 
and  the $2^+_{\rm ms}$ state, are both of the order of $(1/N)^0$, 
i.e., they can be strong on an absolute scale and comparable. 
The $3^-_1 \to 2^+_{\rm ms}$ $F$-vector $E1$ strength can be even 
larger than that of the strong $F$-scalar $3^-_1 \to 2^+_1$ $E1$ 
transition if the effective electric dipole charges $\alpha_\pi$ 
and $\alpha_\nu$ have opposite signs. 
Similarly, the $E1$ ground state excitation strengths to the $1^-_1$ state 
and the $1^-_{\rm ms}$ state are both of the order of $(1/N)^0$ due 
to the two-body part of the $E1$ operator. 
However, here the $F$-scalar $0^+_1 \to 1^-_1$ $E1$ strength must 
be expected to be larger than the $F$-vector $E1$ strength of the 
$0^+_1 \to 1^-_{\rm ms}$ transition because the parameter $\eta$ usually 
takes values between 0 and 1 leading to a reduction of the 
$F$-vector $E1$ strength which is proportional to the factor $(1-\eta)^2$. 
It is also interesting to compare the properties of the new 
mixed-symmetry $1^-$ states, $1^-_{\rm ms}$ and $1^{-\prime}_{\rm ms}$.
The direct $E1$ ground state excitation strengths to the 
$1^{-\prime}_{\rm ms}$ state is only of the order of $(1/N)^2$ and 
therefore much smaller than the $E1$ strength to the $1^-_{\rm ms}$ state, 
which is of the order of $(1/N)^0$. 
The $1^-_{\rm ms} \to 3^-_1$ $E2$ strength is of the order of N, 
which indicates quadrupole collectivity. 
Since it is proportional to the square of the difference 
of the effective quadrupole boson charges $(e_\pi -e_\nu)^2$ 
this transition should be weakly collective like the 
known $2^+_{\rm ms} \to 0^+_1$ one-quadrupole phonon decay. 
In contrast, the $1^{-\prime}_{\rm ms} \to 3^-_1$ $E2$ strength is of the 
order of $1/N$ and cannot be large. 
Very interesting are also the $M1$ transitions between 
symmetric and mixed-symmetry negative parity states. 
The $M1$ strengths from the $1^-_{\rm ms}$ state to the quadrupole-octupole 
coupled $1^-$ and $2^-$ two-phonon states are of the order of 
$(1/N)^0$ and should be comparable to the $2^+_{\rm ms}\to 2^+_1$ 
strong $M1$ transition. 
The $1^{-\prime}_{\rm ms} \to 1^-, 2^-$ $M1$ transitions are 
of the order of $(1/N)^2$ if the $d$-boson part alone is used 
in the $M1$ transition operator (\ref{TM1}). 
If one allows for finite $g$-factors $g^\prime_\rho$ for the $f$-boson 
parts in the $M1$ operator (\ref{TM1}), they will be responsible for 
the leading term of the order of $(1/N)^0$ which was therefore 
included in Table III. 
This resembles the fact that the $1^{-\prime}_{\rm ms}$ state originates 
from the non symmetric coupling of an $f$-boson to the $sd$-sector. 
In summary the $1^-_{\rm ms}$ state shows the behavior, which is 
mentioned in the introduction to be expected for a two-phonon state formed 
by the coupling of the fundamental phonons which generate the 
$3^-_1$ state and the $2^+_{\rm ms}$ state.

\subsubsection{The SO$_{\pi \nu }$(6)$\otimes $U$_{\pi \nu }$(7) 
dynamical symmetry limit}

Using the wave functions (\ref{wfSO6}) we have calculated the matrix elements 
of the transition operators (\ref{TE2}), (\ref{TM1}), (\ref{TE3}) and 
(\ref{TE1}). 
We use here the quadrupole operator (\ref{Q}) with $\chi_{\rho }=0$ as is
conventional for the SO(6) limit.
The analytical expressions for the $E2$ and $M1$ transitions 
between the states which are formed only from $s$ and $d$ bosons 
are identical to those of the $sd$-IBM-2 and can be found in 
Ref.~\cite{PVI86}. 
Analytical expressions for $E2$ and $M1$ transition strengths between 
negative parity states and for $E1$ and $E3$ transition strengths 
are summarized in Table IV.

\section{Summary}

The relation and coupling of the collective octupole degree of freedom to 
proton-neutron mixed-symmetry states is discussed in the framework 
of the interacting boson model. 
For this purpose $s$-bosons, $d$-bosons, and $f$-bosons as well as 
the proton-neutron degree of freedom had to be taken into account 
leading to the hitherto not considered $sdf$-IBM-2 version of the model. 
We developed the formalism of the $sdf$-IBM-2 considering two $F$-spin 
symmetric dynamical symmetry limits which are denoted by 
U$_{\pi \nu }$(5)$\otimes $U$_{\pi \nu }$(7) and 
SO$_{\pi \nu }$(6)$\otimes $U$_{\pi \nu }$(7). 
These analytically solvable limits of the model can be 
relevant for the description of near spherical, vibrational and 
$\gamma $-unstable nuclei, respectively. 
The full dynamically symmetric Hamiltonians are formulated, 
their eigenstates are specified and the analytical expressions for 
their excitation energies are given. 
Low-lying, collective,  negative parity states are discussed 
including for the first time negative parity mixed-symmetry states. 
Analytical expressions for $E1$, $M1$, $E2$, and $E3$ transition strengths 
were evaluated. 
The $E1$ operator considered here contains a quadrupole-octupole 
coupled two-body part besides the one-body term. 
Evidence for the possible dominance of $F$-vector $E1$ transitions 
over $F$-scalar $E1$ transitions is found. 
The decay properties of $1^-_{\rm ms}$ states in vibrational 
nuclei are predicted.

\acknowledgements{We thank C. Fransen for informing us about 
 transition strengths in $^{94}$Mo prior to publication. 
 We gratefully acknowledge discussions with P. von Brentano, R.~F. Casten, 
 C. Fransen, A. Gelberg, F.~Iachello, J. Jolie, R.~V. Jolos, 
 T.~Otsuka, A.~Wolf, N.V. Zamfir, and A. Zilges. 
 N.A.S., N.P. and P.V.I. thank the Institute for Nuclear Theory, 
 University of Washington for hospitality.
 This work was partially supported by the Competitive Center at 
 the St.~Petersburg State University, 
 by Grant-in-Aid for Scientific Research (B)(2)(10044059) from 
 the Ministry of Education, Science and Culture, 
 by the U.S. DOE under Contract No. DE-FG02-91ER-40609, 
 and by the Deutsche Forschungsgemeinschaft under Contracts 
 No. Br 799/9-1 and Pi 393/1-1.}

\begin{appendix}

\section{U(13)$\supset $ U(6)$\otimes $U(7) branching rules} 

In order to provide a complete classification scheme we need to reduce the
representations of U(13) group to the Kronecker product U(6)$\otimes $U(7). 
Here we present the branching rules only for the totally symmetric $[N]$ and 
the lowest mixed-symmetry $[N-1,1]$ and $[N-2,2]$ irreducible 
representations of U(13). 
The  U(6)$\otimes $U(7) representations are denoted 
as $[\alpha ]\otimes [\beta ]$. 
Those U(6)$\otimes $U(7) decompositions, which 
contain $1^-$ states discussed in the text, are underlined.  
\begin{equation}
\begin{array}{ll}
[N]=&[N]\otimes[0], \; \underline{[N-1]\otimes[1]}, \ldots , [1]\otimes[N-1], 
\;   [0]\otimes[N] \\[2pt]    
[N-1,1]=&[N-1,1]\otimes[0], \; \underline{[N-2,1]\otimes[1]}, \ldots , 
         [2,1]\otimes[N-3], 
\;       [1,1]\otimes[N-2] \\        
        &\underline{[N-1]\otimes[1]}, \; [N-2]\otimes[2], \ldots , 
         [2]\otimes[N-2], 
\;       [1]\otimes[N-1] \\   
        &[N-2]\otimes[1,1], \; [N-3]\otimes[2,1], \ldots , [1]\otimes[N-2,1], 
\;       [0]\otimes[N-1,1] \\[2pt]         
[N-2,2]=&[N-2,2]\otimes[0], \; [N-3,2]\otimes[1], \ldots , [3,2]\otimes[N-5], 
\;       [2,2]\otimes[N-4] \\  
        &\underline{[N-2,1]\otimes[1]}, \; [N-3,1]\otimes[2], \ldots , 
         [3,1]\otimes[N-4], 
\;       [2,1]\otimes[N-3] \\
        &[N-2]\otimes[2], \; [N-3]\otimes[3], \ldots , [3]\otimes[N-3], 
\;       [2]\otimes[N-2] \\
        &[N-3,1]\otimes[1,1], \; [N-4,1]\otimes[2,1], \ldots , 
\;       [2,1]\otimes[N-4,1], \; [1,1]\otimes[N-3,1] \\
         &[N-3]\otimes[2,1], \; [N-4]\otimes[3,1], \ldots , [2]\otimes[N-3,1],
\;       [1]\otimes[N-2,1] \\
        &[N-4]\otimes[2,2], \; [N-5]\otimes[3,2], \ldots , [1]\otimes[N-3,2], 
\;       [0]\otimes[N-2,2] \\ [2pt] 
\end{array}
\end{equation}  
The branching rules for the chains  
U(6)$\supset $ U(5)$\supset $ SO(5) $\supset $SO(3) and
U(6)$\supset $ SO(6)$\supset $ SO(5) $\supset $SO(3)  reduction chains
can be found elsewhere~\cite{IaAr87}.
The partial classification scheme for the chain
U(7)$\supset $ SO(7)$\supset $G$_2$ $\supset $SO(3) 
is given in Table VII.

\section{Casimir invariants}

The Casimir invariants of Lie algebras exploited here can be expressed
in terms of boson creation and annihilation operators as follows~\cite{IaAr87}:
\begin{equation}
\label{Cas_inv}
\begin{array}{ll}
%C_1\left[\mbox{U(13)}\right]=& \left(s^{+}\cdot s\right) 
%+\left(d^{+}\cdot \tilde d\right)+\left(f^{+}\cdot \tilde f\right) \;, \\
%C_2\left[\mbox{U(13)}\right]=& (s^{+} \cdot s) \cdot (s^{+}\cdot s) 
%+\sum\limits_{K=0}^4 \left[d^{+} \times \tilde d\right]^{(K)}\cdot 
%\left[d^{+} \times \tilde d\right]^{(K)} 
%+\sum\limits_{K=0}^6 \left[f^{+} \times \tilde f\right]^{(K)}\cdot 
%\left[f^{+} \times \tilde f\right]^{(K)} \;, \\ 

C_1\left[\mbox{U(7)}\right]=&\left(f^{+}\cdot \tilde f\right) \;, \\
C_2\left[\mbox{U(7)}\right]=&\sum\limits_{K=0}^6 
\left[f^{+} \times \tilde f\right]^{(K)}\cdot 
\left[f^{+} \times \tilde f\right]^{(K)} \;, \\ 

C_1\left[\mbox{U(6)}\right]=&\left(s^{+}\cdot s\right)
+\left(d^{+}\cdot \tilde d\right) \;, \\
C_2\left[\mbox{U(6)}\right]=& (s^{+} \cdot s) \cdot (s^{+}\cdot s) 
+\sum\limits_{K=0}^4 \left[d^{+} \times \tilde d\right]^{(K)}\cdot 
\left[d^{+} \times \tilde d\right]^{(K)} \;, \\ 

C_2\left[\mbox{SU(3)}\right]=&
2 \left(\left[d^{+}{\times }s +s^+{\times }\tilde d \right]^{(2)} 
-\frac{\sqrt{7}}{2}\left[d^{+}{\times }\tilde d \right]^{(2)}\right)
\left(\left[d^{+}{\times }s +s^+{\times }\tilde d \right]^{(2)} 
-\frac{\sqrt{7}}{2}\left[d^{+}{\times }\tilde d \right]^{(2)}\right)\\
&+\frac{15}{2}\left[d^{+}{\times }\tilde d \right]^{(1)}\cdot 
\left[d^{+}{\times }\tilde d \right]^{(1)} \; , \\

C_2\left[\mbox{SO(7)}\right]=&\sum\limits_{K=1,3,5,7}
\left[f^{+} \times \tilde f\right]^{(K)}\cdot 
\left[f^{+} \times \tilde f\right]^{(K)} \; ,  \\

C_2\left[\mbox{SO(6)}\right]=&
\left[d^{+}{\times }s +s^+{\times }\tilde d \right]^{(2)} \cdot
\left[d^{+}{\times }s +s^+{\times }\tilde d \right]^{(2)}    
+2 \sum\limits_{K=1,3} \left[d^{+}{\times }\tilde d \right]^{(K)}\cdot 
\left[d^{+}{\times }\tilde d \right]^{(K)} \; , \\

C_2\left[\mbox{SO(5)}\right]=&\sum\limits_{K=1,3}
\left[d^{+} \times \tilde d\right]^{(K)}\cdot 
\left[d^{+} \times \tilde d\right]^{(K)} \; ,  \\

C_2\left[\mbox{SO$^d$(3)}\right]=&
10\left[d^{+} \times \tilde d\right]^{(1)}\cdot 
\left[d^{+} \times \tilde d\right]^{(1)} \; ,  \\

C_2\left[\mbox{SO$^f$(3)}\right]=&
28\left[f^{+} \times \tilde f\right]^{(1)}\cdot 
\left[f^{+} \times \tilde f\right]^{(1)} \; ,  \\

C_2\left[\mbox{G}_2\right]=&\sum\limits_{K=1,5}
\left[f^{+}{\times }\tilde f \right]^{(K)}\cdot 
\left[f^{+}{\times }\tilde f \right]^{(K)} \; . \\                
\end{array}
\end{equation}   
The exceptional algebra G$_2$ has been introduced in physics 
by Racah~\cite{Racah}. 
The eigenvalues of the Casimir invariants from (\ref{Cas_inv})
in an irreducible representation 
given by a $[f_1 f_2 \ldots f_n]$ Young tableau are 
\begin{equation}
\label{Cas_eigen}
\begin{array}{l}
\langle C_1\left[\mbox{U}(n)\right]\rangle =\sum\limits_{i=1}^{n} f_i \;, \\
\langle C_2\left[\mbox{U}(n)\right]\rangle =\sum\limits_{i=1}^{n} 
f_i (f_i+n+1-2 i) \;, \\              
\langle C_2\left[\mbox{SU(3)}\right]\rangle =\lambda^2 + \mu^2 +\lambda \mu
+3 \lambda +3 \mu \; , \qquad (\lambda=f_1-f_2, \mu = f_2) \\              
\langle C_2\left[\mbox{SO}(2n+1)\right]\rangle = \sum\limits_{i=1}^{n} 
2 f_i (f_i+2 n+1-2 i) \;, \\   
\langle C_2\left[\mbox{SO}(2n)\right]\rangle = \sum\limits_{i=1}^{n} 
2 f_i (f_i+2 n-2 i) \;, \\                    
\langle C_2\left[\mbox{G}_2\right]\rangle =f_1(f_1+5)+f_2(f_2 +4)+f_1f_2 \; .   
\end{array}
\end{equation}

\end{appendix}

\begin{table}
\label{tab:WvfctsU5}     
\caption{Notation for some low-lying states in the  
U$_{\pi \nu }$(1)$\otimes $U$_{\pi \nu }$(5)$\otimes $U$_{\pi \nu }$(7) 
dynamical symmetry limit. }
\begin{tabular}{ll}
& ${{|[N_{\pi }] {\otimes }[N_{\nu }];[N_1,N_2],[n_1,n_2]\{n_{d_1},n_{d_2}\}
(\tau_1,\tau_2)\{\alpha_i\} L_d;[m_1,m_2](\omega_1,\omega_2)(u_1,u_2)
\{\beta_j\} L_f;L 
\rangle }}$ 
\\[3pt]
\hline                                                       
$|0^+_1 \rangle $ & = 
$|[N_{\pi }] \otimes [N_{\nu }];[N],[N]\{0\}(0)0;[0](0)(0)0;0 \rangle  
= |s_{\pi }^{N_{\pi }} s_{\nu }^{N_{\nu }};0 \rangle $ \\[3pt]

$|0^+_2 \rangle $ & 
$=|[N_{\pi }] \otimes [N_{\nu }];[N],[N]\{2\}(0)0;[0](0)(0)0;0 \rangle $ \\
& $= (N(N-1))^{-1/2} 
\left(\sqrt{N_{\pi }(N_{\pi }-1)}|d_{\pi }^2 ;0 \rangle    
+\sqrt{2N_{\nu }N_{\pi }}|d_{\pi } d_{\nu };0 \rangle    
+\sqrt{N_{\nu }(N_{\nu }-1)}|d_{\nu }^2;0 \rangle \right) $ \\[3pt]

$|1^+_{\rm ms} \rangle $ & 
$=|[N_{\pi }] \otimes [N_{\nu }];[N-1,1],[N-1,1]\{1,1\}(1,1)1;[0](0)(0)0;1 
\rangle $ \\
& $=|d_{\pi }d_{\nu };1 \rangle  $ \\[3pt]

$|2^+_1 \rangle $ & 
$=|[N_{\pi }] \otimes [N_{\nu }];[N],[N]\{1\}(1)2;[0](0)(0)0;2 \rangle $ \\
& $= N^{-1/2} 
\left(\sqrt{N_{\pi }}|d_{\pi };2 \rangle    
+\sqrt{N_{\nu }}|d_{\nu };2 \rangle \right) $ \\[3pt]
 
$|2^+_2 \rangle $ & 
$=|[N_{\pi }] \otimes [N_{\nu }];[N],[N]\{2\}(2)2;[0](0)(0)0;2 \rangle $ \\
& $=(N(N-1))^{-1/2} 
\left(\sqrt{N_{\pi }(N_{\pi }-1)}|d_{\pi }^2 ;2 \rangle    
+\sqrt{2N_{\nu }N_{\pi }}|d_{\pi } d_{\nu };2 \rangle    
+\sqrt{N_{\nu }(N_{\nu }-1)}|d_{\nu }^2;2 \rangle \right) $ \\[3pt]
  
$|2^+_{\rm ms} \rangle $ & 
$=|[N_{\pi }] \otimes [N_{\nu }];[N-1,1],[N-1,1]\{1\}(1)2;[0](0)(0)0;2 
\rangle $ \\
& $=N^{-1/2}\left(\sqrt{N_{\nu }}|d_{\pi };2 \rangle    
-\sqrt{N_{\pi }}|d_{\nu };2 \rangle \right) $ \\[3pt] 
 
$|3^+_1 \rangle $ & 
$=|[N_{\pi }] \otimes [N_{\nu }];[N],[N]\{3\}(3)3;[0](0)(0)0;3 \rangle $ \\
& $= (N(N-1)(N-2))^{-1/2} 
\left(\sqrt{N_{\pi }(N_{\pi }-1)(N_{\pi }-2)}|d_{\pi }^3 ;3 \rangle    
+\sqrt{N_{\nu }(N_{\nu }-1)(N_{\nu }-2)}|d_{\nu }^3;3 \rangle  \right.$ \\
& $ +\sqrt{3N_{\pi }(N_{\pi }-1)N_{\nu }}
\left(\sqrt{\frac57}|d_{\pi } (2) d_{\nu };3 \rangle    
-\sqrt{\frac27}|d_{\pi } (4) d_{\nu };3 \rangle \right) $ \\  
& $ \left. +\sqrt{3N_{\pi }N_{\nu }(N_{\nu }-1)}
\left(\sqrt{\frac57}|d_{\nu } (2) d_{\pi };3 \rangle    
-\sqrt{\frac27}|d_{\nu } (4) d_{\pi };3 \rangle \right) \right) $ \\[3pt]

$|3^+_{\rm ms} \rangle $ & 
$=|[N_{\pi }] \otimes [N_{\nu }];[N-1,1],[N-1,1]\{1,1\}(1,1)3;[0](0)(0)0;3 
\rangle $ \\
& $=|d_{\pi }d_{\nu };3 \rangle  $ \\[3pt]

$|4^+_1 \rangle $ & 
$=|[N_{\pi }] \otimes [N_{\nu }];[N],[N]\{2\}(2)4;[0](0)(0)0;4 \rangle $\\
& $= (N(N-1))^{-1/2} 
\left(\sqrt{N_{\pi }(N_{\pi }-1)}|d_{\pi }^2 ;4 \rangle    
+\sqrt{2N_{\nu }N_{\pi }}|d_{\pi } d_{\nu };4 \rangle    
+\sqrt{N_{\nu }(N_{\nu }-1)}|d_{\nu }^2;4 \rangle \right) $ \\[3pt]

$|6^+_1 \rangle $ & 
$=|[N_{\pi }] \otimes [N_{\nu }];[N],[N]\{3\}(3)6;[0](0)(0)0;6 \rangle $ \\
& $= (N(N-1)(N-2))^{-1/2} 
\left(\sqrt{N_{\pi }(N_{\pi }-1)(N_{\pi }-2)}|d_{\pi }^3 ;6 \rangle    
+\sqrt{N_{\nu }(N_{\nu }-1)(N_{\nu }-2)}|d_{\nu }^3;6 \rangle  \right.$ \\
& $ +\sqrt{3N_{\pi }(N_{\pi }-1)N_{\nu }}
\left(\sqrt{\frac57}|d_{\pi } (2) d_{\nu };6 \rangle    
-\sqrt{\frac27}|d_{\pi } (4) d_{\nu };6 \rangle \right) $ \\  
& $ \left. +\sqrt{3N_{\pi }N_{\nu }(N_{\nu }-1)}
\left(\sqrt{\frac57}|d_{\nu } (2) d_{\pi };6 \rangle    
-\sqrt{\frac27}|d_{\nu } (4) d_{\pi };6 \rangle \right) \right) $ \\[3pt]

$|3^-_1 \rangle $ & 
$=|[N_{\pi }] \otimes [N_{\nu }];[N],[N-1]\{0\}(0)0;[1](1)(1)3;3\rangle $ \\
& $= N^{-1/2} 
\left(\sqrt{N_{\pi }}|f_{\pi };3 \rangle    
+\sqrt{N_{\nu }}|f_{\nu };3 \rangle \right) $ \\[3pt]

$|3^-_{\rm ms} \rangle $ & 
$=|[N_{\pi }] \otimes [N_{\nu }];[N-1,1],[N-1]\{0\}(0)0;[1](1)(1)3;3
\rangle $ \\
& $= N^{-1/2} \left(\sqrt{N_{\nu }}|f_{\pi };3 \rangle    
-\sqrt{N_{\pi }}|f_{\nu };3 \rangle \right) $ \\[3pt]

$|1^-_1 \rangle $ & 
$=|[N_{\pi }] \otimes [N_{\nu }];[N],[N-1]\{1\}(1)2;[1](1)(1)3;1\rangle $ \\
& $= (N(N-1))^{-1/2} 
\left(\sqrt{N_{\pi }(N_{\pi }-1)}|d_{\pi } f_{\pi } ;1 \rangle 
+\sqrt{N_{\nu }(N_{\nu }-1)}|d_{\nu } f_{\nu } ;1 \rangle \right.$ \\
& $\left. +\sqrt{N_{\nu }N_{\pi }}|d_{\pi } f_{\nu };1 \rangle    
+\sqrt{N_{\nu }N_{\pi }}|d_{\nu } f_{\pi };1 \rangle \right) $  \\[3pt]

$|1^{-}_{\rm ms} \rangle $ & 
$=|[N_{\pi }] \otimes [N_{\nu }];[N-1,1],[N-2,1]\{1\}(1)2;[1](1)(1)3;1 
\rangle $ \\  
& $=((N-1)(N-2))^{-1/2} 
\left(\sqrt{N_{\nu }(N_{\pi }-1)}|d_{\pi } f_{\pi } ;1 \rangle 
-\sqrt{N_{\pi }(N_{\nu }-1)}|d_{\nu } f_{\nu } ;1 \rangle \right. $ \\
& $ \left. +(N_{\nu }-1) |d_{\pi } f_{\nu };1 \rangle    
-(N_{\pi }-1) |d_{\nu } f_{\pi };1 \rangle \right)  $ \\[3pt]

$|1^{- \prime }_{\rm ms} \rangle $ & 
$=|[N_{\pi }] \otimes [N_{\nu }];[N-1,1],[N-1]\{1\}(1)2;[1](1)(1)3;1 
\rangle $ \\  
& $= (N(N-1))^{-1/2} 
\left(\sqrt{N_{\nu }(N_{\pi }-1)}|d_{\pi } f_{\pi } ;1 \rangle 
-\sqrt{N_{\pi }(N_{\nu }-1)}|d_{\nu } f_{\nu } ;1 \rangle \right. $ \\
& $\left. -N_{\pi } |d_{\pi } f_{\nu };1 \rangle    
+N_{\nu } |d_{\nu } f_{\pi };1 \rangle \right) $  \\[3pt]

$|1^{- \prime \prime }_{\rm ms} \rangle $ &
$=|[N_{\pi }] \otimes [N_{\nu }];[N-2,2],[N-2,1]\{1\}(1)2;[1](1)(1)3;1 
\rangle $ \\  
& $=((N-1)(N-2))^{-1/2} 
\left(-\sqrt{N_{\nu }(N_{\nu }-1)}|d_{\pi } f_{\pi } ;1 \rangle 
-\sqrt{N_{\pi }(N_{\pi }-1)}|d_{\nu } f_{\nu } ;1 \rangle \right. $ \\
& $\left. +\sqrt{(N_{\pi }-1)(N_{\nu }-1)} |d_{\pi } f_{\nu };1 \rangle    
+\sqrt{(N_{\pi }-1)(N_{\nu }-1)} |d_{\nu } f_{\pi };1 \rangle \right)$\\[3pt]

$|2^-_1 \rangle $ & 
$=|[N_{\pi }] \otimes [N_{\nu }];[N],[N-1]\{1\}(1)2;[1](1)(1)3;2 \rangle $ \\
& $= (N(N-1))^{-1/2} 
\left(\sqrt{N_{\pi }(N_{\pi }-1)}|d_{\pi } f_{\pi } ;2 \rangle 
+\sqrt{N_{\nu }(N_{\nu }-1)}|d_{\nu } f_{\nu } ;2 \rangle \right. $ \\
& $\left. +\sqrt{N_{\nu }N_{\pi }}|d_{\pi } f_{\nu };2 \rangle    
+\sqrt{N_{\nu }N_{\pi }}|d_{\nu } f_{\pi };2 \rangle \right) $ \\[3pt]

$|3^-_2 \rangle $ & = 
$|[N_{\pi }] \otimes [N_{\nu }];[N],[N-1]\{1\}(1)2;[1](1)(1)3;3 \rangle $ \\
& $= (N(N-1))^{-1/2} 
\left(\sqrt{N_{\pi }(N_{\pi }-1)}|d_{\pi } f_{\pi } ;3 \rangle 
+\sqrt{N_{\nu }(N_{\nu }-1)}|d_{\nu } f_{\nu } ;3 \rangle \right. $ \\
& $\left. +\sqrt{N_{\nu }N_{\pi }}|d_{\pi } f_{\nu };3 \rangle    
+\sqrt{N_{\nu }N_{\pi }}|d_{\nu } f_{\pi };3 \rangle \right) $ \\[3pt]

$|4^-_1 \rangle $ & 
$=|[N_{\pi }] \otimes [N_{\nu }];[N],[N-1]\{1\}(1)2;[1](1)(1)3;4 
\rangle $ \\
& $= (N(N-1))^{-1/2} 
\left(\sqrt{N_{\pi }(N_{\pi }-1)}|d_{\pi } f_{\pi } ;4 \rangle 
+\sqrt{N_{\nu }(N_{\nu }-1)}|d_{\nu } f_{\nu } ;4 \rangle \right. $ \\
& $ \left. +\sqrt{N_{\nu }N_{\pi }}|d_{\pi } f_{\nu };4 \rangle    
+\sqrt{N_{\nu }N_{\pi }}|d_{\nu } f_{\pi };4 \rangle \right) $ \\[3pt]

$|5^-_1 \rangle $ & 
$=|[N_{\pi }] \otimes [N_{\nu }];[N],[N-1]\{1\}(1)2;[1](1)(1)3;5\rangle $ \\ 
& $= (N(N-1))^{-1/2} 
\left(\sqrt{N_{\pi }(N_{\pi }-1)}|d_{\pi } f_{\pi } ;5 \rangle 
+\sqrt{N_{\nu }(N_{\nu }-1)}|d_{\nu } f_{\nu } ;5 \rangle \right. $ \\
& $\left. +\sqrt{N_{\nu }N_{\pi }}|d_{\pi } f_{\nu };5 \rangle    
+\sqrt{N_{\nu }N_{\pi }}|d_{\nu } f_{\pi };5 \rangle \right) $ \\
\end{tabular}
\end{table}

\begin{table}
\label{tab:WvfctsSO6}     
\caption{Notation for some low-lying states in the 
SO$_{\pi \nu }$(6)$\otimes $U$_{\pi \nu }$(7) dynamical symmetry limit. }
\begin{tabular}{l}

${{|[N_{\pi }] {\otimes }[N_{\nu }];[N_1,N_2],[n_1,n_2] 
\langle \sigma_1,\sigma_2 \rangle (\tau_1,\tau_2)\{\alpha_i \} L_d;
[m_1,m_2](\omega_1,\omega_2)(u_1,u_2)\{\beta_j\} L_f;LM \rangle  }}$ \\ 
\hline                                                       
$|0^+_1 \rangle = 
|[N_{\pi }] \otimes [N_{\nu }];[N],[N]\langle N \rangle (0)0;[0](0)(0)0;0 
\rangle $ \\[3pt]
 
$|0^+_2 \rangle = 
|[N_{\pi }] \otimes [N_{\nu }];[N],[N] \langle N \rangle (2)0;[0](0)(0)0;0 
\rangle $ \\[3pt]

$|1^+_{\rm ms} \rangle = 
|[N_{\pi }] \otimes [N_{\nu }];[N-1,1],[N-1,1]\langle N-1,1 \rangle (1,1) 1;
[0](0)(0)0;1 \rangle $ \\[3pt]
    
$|2^+_1 \rangle = 
|[N_{\pi }] \otimes [N_{\nu }];[N],[N]\langle N \rangle (1)2;[0](0)(0)0;2 
\rangle $ \\[3pt]
 
$|2^+_2 \rangle = 
|[N_{\pi }] \otimes [N_{\nu }];[N],[N] \langle N \rangle (2)2;[0](0)(0)0;2 
\rangle $ \\[3pt]
  
$|2^+_{\rm ms} \rangle = 
|[N_{\pi }] \otimes [N_{\nu }];[N-1,1],[N-1,1]\langle N-1,1 \rangle (1)2;
[0](0)(0)0;2 \rangle $ \\[3pt]

$|3^+_{\rm ms} \rangle = 
|[N_{\pi }] \otimes [N_{\nu }];[N-1,1],[N-1,1]\langle N-1,1 \rangle (1,1) 3;
[0](0)(0)0;3 \rangle $ \\[3pt]

$|4^+_1 \rangle = 
|[N_{\pi }] \otimes [N_{\nu }];[N],[N]\langle 2 \rangle (2)4;[0](0)(0)0;4 
\rangle $ \\[3pt]
  
$|3^-_1 \rangle = 
|[N_{\pi }] \otimes [N_{\nu }];[N],[N-1]\langle N-1 \rangle (0)0;[1](1)(1)3;3 
\rangle $ \\[3pt]

$|3^{- \prime }_{\rm ms} \rangle = 
|[N_{\pi }] \otimes [N_{\nu }];[N-1,1],[N-1] \langle N-1 \rangle (0)0;
[1](1)(1)3;3 \rangle $ \\[3pt]

$|1^-_1 \rangle = 
|[N_{\pi }] \otimes [N_{\nu }];[N],[N-1] \langle N-1 \rangle (1)2;[1](1)(1)3;1 
\rangle $ \\[3pt]  

$|1^-_{\rm ms} \rangle = 
|[N_{\pi }] \otimes [N_{\nu }];[N-1,1],[N-2,1] \langle N-2,1 \rangle (1)2;
[1](1)(1)3;1 \rangle $ \\[3pt]

$|1^{- \prime }_{\rm ms} \rangle = 
|[N_{\pi }] \otimes [N_{\nu }];[N-1,1],[N-1] \langle N-1 \rangle (1)2;
[1](1)(1)3;1 \rangle $ \\[3pt]

$|1^{- \prime \prime }_{\rm ms} \rangle = 
|[N_{\pi }] \otimes [N_{\nu }];[N-2,2],[N-2,1] \langle N-2,1 \rangle (1)2;
[1](1)(1)3;1 \rangle $ \\  
\end{tabular}
\end{table}

\begin{table}
\label{tab:emU5} 
\caption{Some analytical expressions for reduced electromagnetic 
transition strengths in the  
U$_{\pi \nu }$(1)$\otimes $U$_{\pi \nu }$(5)$\otimes $U$_{\pi \nu }$(7) 
dynamical symmetry limit. }
\begin{tabular}{lcl}

$B(E3;0^{+}_1 \to 3^{-}_1)$&=&$7 e_3^2 N $  \\[5pt]
\hline
$B(E1;3^{-}_1 \to 2^{+}_1)$&=&$\frac37 
\left[(\alpha_{\nu } N_{\nu } + \alpha_{\pi } N_{\pi })\frac1N+
\beta \left(N_{\pi }+\eta N_{\nu }\right) \frac{2\,N-1}{2N^2} + 
\sqrt{{6\over {35}}} \beta \chi \left(\chi_{\pi } N_{\pi } 
+ \chi_{\nu } \eta N_{\nu }\right)\frac{1}{N^2}\right]^2$\\[5pt]

$B(E1;3^{-}_1\to 2^{+}_{\rm ms})$&=&$\frac37\left[\alpha_{\pi }-\alpha_{\nu }+ 
\beta \, \left(1 - \eta \right)\frac{2\,N-1}{2N}  
+ \sqrt{{6\over {35}}} \beta \,\chi \,
\left(\chi_{\pi  } -\eta \chi_{\nu } \right)\frac1N \, \right]^2
\frac{N_{\nu } \,N_{\pi }}{N^2}  $ \\[5pt]         

$B(E1;1^{-}_1 \to 0^{+}_1)$&=&$\beta^2 
\left(N_{\pi }+ \eta N_{\nu } \right)^2 \frac{N-1}{N^3} $ \\[5pt]

$B(E1;1^{-}_1 \to 2^{+}_1)$&=&$\frac{6}{25}\,{\beta ^2}\,
\left[ \chi \,\left(N_{\pi } + \eta N_{\nu } \right)  + 
\sqrt{\frac{5}{42}}\,
\left(\chi_{\pi  } \, N_{\pi } + \eta \chi_{\nu }\,N_{\nu }\right)\right]^2
\frac{N-1}{N^4}$ \\[5pt]

$B(E1;1^{-}_1 \to 0^{+}_2)$&=&$\frac25 
\left[\left(\alpha_{\nu }N_{\nu }+\alpha_{\pi }N_{\pi}\right)\frac1N + 
\beta \left(N_{\pi }+\eta N_{\nu }\right) \frac{2\,N-3}{2N^2} +
3\,\sqrt{6\over {35}} \beta \chi \, \left(\chi_{\pi  } N_{\pi }+ 
\eta \chi_{\nu } N_{\nu } \right)\frac{1}{N^2} \right]^2 $ \\[5pt]

$B(E1;1^{-}_1 \to 2^{+}_2)$&=&$\frac{2}{35}
\left[\left(\alpha_{\nu }N_{\nu }+\alpha_{\pi }N_{\pi } \right) \frac1N
- \beta \,\left(N_{\pi }+ \eta N_{\nu } \right)\frac{2 N-3}{2N^2}
-\sqrt{\frac{6}{35}} \,\beta \, \chi \,
\left(\chi_{\pi  } N_{\pi }+ \eta \chi_{\nu }N_{\nu }\right]\frac{1}{N^2}
\right]^2 $ \\[5pt]

$B(E1;4^{-}_1 \to 4^{+}_1)$&=&$\frac17 
\left[\left(\alpha_{\nu }N_{\nu }+ \alpha_{\pi }N_{\pi }\right)\frac1N + 
\beta \,\left(N_{\pi }+\eta N_{\nu }\right)\frac{2N -3}{2N^2}+ 
3\, \sqrt{{6\over {35}}} \,\beta \chi \left(\chi_{\pi  }N_{\pi }
+\eta \chi_{\nu }N_{\nu } \right) \frac{1}{N^2} \right]^2$ \\[5pt]

$B(E1;5^{-}_1 \to 4^{+}_1)$&=&$\frac67 
\left[\left(\alpha_{\nu }N_{\nu }+\alpha_{\pi }N_{\pi }\right)\frac1N + 
\beta \left(N_{\pi }+\eta N_{\nu }\right)\frac{2N -3}{2 N^2}+ 
4\,\sqrt{{2\over {105}}} \beta \chi 
\left(\chi_{\pi  }N_{\pi }+\eta \chi_{\nu } N_{\nu }\right)\frac{1}{N^2}
\right]^2 $ \\[5pt]

$B(E1;3^{-}_1 \to 4^{+}_1)$&=&$\frac{3}{245}\,{\beta^2} {\chi }^2 
\left(N_{\pi }+\eta N_{\nu }\right)^2 \frac{N -1}{N^4} $ \\[5pt]

$B(E1;3^{-}_1 \to 2^{+}_2)$&=&$\frac{144}{245}\,{\beta^2} {\chi }^2 
\left(N_{\pi }+\eta N_{\nu }\right)^2 \frac{N -1}{N^4} $ \\[5pt]

$B(E1;4^{-}_1 \to 3^{+}_1)$&=&$\frac{289}{280}\,{\beta^2} {\chi }^2 
\left(N_{\pi }+\eta N_{\nu }\right)^2 \frac{N -2}{N^4} $ \\[5pt]

$B(E1;5^{-}_1 \to 6^{+}_1)$&=&$\frac{13}{385}\,{\beta^2} {\chi }^2 
\left(N_{\pi }+\eta N_{\nu }\right)^2 \frac{N -2}{N^4} $ \\[5pt]  

$B(E2;1^{-  }_{\rm ms} \to 3^{-}_1)$&=&
$\left(e_{\pi }- e_{\nu }\right)^2  
\frac{N_{\pi } \,N_{\nu } (N-2)}{N(N-1)}  $ \\[5pt]

$B(M1;1^{-  }_{\rm ms} \to 1^{-}_1)$&=&$\frac{3}{2 \pi } 
\left(g_{\pi }- g_{\nu }\right)^2  
\frac{N_{\pi } \,N_{\nu } (N-2)}{N(N-1)^2}  $ \\[5pt]

$B(M1;1^{-  }_{\rm ms}\to 2^{-}_1)$&=&$\frac{3}{\pi } 
\left(g_{\pi }- g_{\nu }\right)^2  
\frac{N_{\pi } \,N_{\nu } (N-2)}{N(N-1)^2}  $ \\[5pt]

$B(E1;1^{- }_{\rm ms} \to 0^{+}_1)$&=&$\beta^2 (1-\eta )^2  
\frac{N_{\pi } \,N_{\nu } (N-2)}{N^2 (N-1)}$ \\[5pt]

%%* $B(E1;1^{- }_{\rm ms} \to 2^{+}_1)$&=&$\frac{1}{35} \beta^2 
%%\left(\chi_{\pi } - \eta \chi_{\nu }
%%+\sqrt{\frac{42}{5}} (1-\eta )\chi \right)^2  
%%\frac{N_{\pi } \,N_{\nu } (N-2)}{N^3 (N-1)}$ \\[5pt] 
%% 
%% 
%%* $B(E1;1^{- }_{\rm ms} \to 1^{+}_{\rm ms})$&=&$\frac{1}{5} 
%%\left[ \beta (N_{\pi } + \eta N_{\nu }) \frac{2N-5}{2N}
%%-\sqrt{\frac{6}{35}} \beta \chi (\chi_{\pi } + \eta \chi_{\nu }) 
%%\frac{2N-5}{2N} 
%%+\alpha_{\pi } (N_{\pi }-1)+\alpha_{\nu } (N_{\nu }-1) \right. $\\
%%& & $\left. -\frac12 \sqrt{\frac{6}{35}} \beta \chi 
%%(\chi_{\pi } N_{\pi }+ \eta \chi_{\nu }N_{\nu }) \frac1N
%%+\frac12 \beta (1+\eta )\frac1N \right]^2  
%%\frac{1}{(N-1)(N-2)}$ \\[5pt] 
%% 

$B(M1;3^{- }_{\rm ms} \to 3^{-}_1)$&=&$\frac{9}{\pi}
\left(g'_{\pi } - g'_{\nu }\right)^2 \frac{N_{\pi } \,N_{\nu }}{N^2}$ \\[5pt]

$B(E1;3^{- }_{\rm ms} \to 2^{+}_1)$&=&$\frac{3}{7} 
\left[\alpha_{\pi} - \alpha_{\nu} -\frac12 \beta (1-\eta )\frac1N 
+\sqrt{\frac{6}{35}}\beta \chi (\chi_{\pi }-\eta \chi_{\nu })\frac1N \right]^2
\frac{N_{\pi } \,N_{\nu }}{N^2}$ \\[5pt]

$B(E1;3^{- }_{\rm ms} \to 2^{+}_{\rm ms})$&=&$\frac{3}{7} 
\left[(\alpha_{\pi} N_{\pi } + \alpha_{\nu} N_{\nu }) 
-\frac12 \beta (N_{\nu }+\eta N_{\pi })\frac1N
+\sqrt{\frac{6}{35}}\beta \chi (\chi_{\pi }N_{\nu }
+\eta \chi_{\nu }N_{\pi })\frac1N \right]^2
\frac{1}{N^2}$ \\[5pt]

$B(E2;1^{- \prime}_{\rm ms} \to 3^{-}_1)$&=&$ 
\left(e_{\pi }- e_{\nu }\right)^2  
\frac{N_{\pi } \,N_{\nu }}{N^2(N-1)}  $ \\[5pt]

$B(M1;1^{- \prime}_{\rm ms} \to 1^{-}_1)$&=&$\frac{3}{2 \pi } 
\left[\frac{g_{\pi }- g_{\nu }}{N-1}-2 (g'_{\pi }-g'_{\nu })\right]^2  
\frac{N_{\pi } \,N_{\nu }}{N^2}  $ \\[5pt]

$B(M1;1^{- \prime}_{\rm ms}\to 2^{-}_1)$&=&$\frac{3}{\pi } 
\left[\frac{g_{\pi }- g_{\nu }}{N-1}- (g'_{\pi }-g'_{\nu })\right]^2  
\frac{N_{\pi } \,N_{\nu }}{N^2}  $ \\[5pt]

$B(E1;1^{- \prime}_{\rm ms} \to 0^{+}_1)$&=&$\beta^2 (1-\eta )^2  
\frac{N_{\pi } \,N_{\nu }}{N^3 (N-1)}$ \\[5pt]

$B(E1;1^{- \prime}_{\rm ms} \to 2^{+}_1)$&=&$\frac{1}{35} \beta^2 
\left[\chi_{\pi }-\eta \chi_{\nu }
+\sqrt{\frac{42}{5}} (1-\eta )\chi \right]^2   
\frac{N_{\pi } \,N_{\nu }}{N^4 (N-1)}$ \\
%$B(E1;1^{- \prime }_{\rm ms} \to 1^{+}_{\rm ms})$&=&$\frac15 
%\left[\beta \left(N_{\pi } + \eta N_{\nu }\right) \frac{2N-5}{2N} 
%-2\sqrt{\frac{6}{35}} \beta \chi 
%\frac{\left(\chi_{\pi }N_{\pi } + \eta \chi_{\nu }N_{\nu }\right)}{N}
%+\alpha_{\pi }N_{\pi }+\alpha_{\nu }N_{\nu } \right. $ \\
%& & $\left. \frac12 \frac{\beta }{N} (1 + \eta)  
%+\sqrt{\frac{6}{35}} \frac{\beta }{N}\chi 
%\left(\chi_{\pi } + \eta \chi_{\nu }\right)
%\right]^2 \frac{1}{(N-1)(N-2)} $ \\  
\end{tabular} 
\end{table}

\begin{table}
\label{tab:emSO6} 
\caption{Some analytical expressions for reduced electromagnetic 
transition strengths in the SO$_{\pi \nu }$(6)
$\otimes $ U$_{\pi \nu }$(7) dynamical symmetry limit. }
\begin{tabular}{lcl}
$\displaystyle B(E3;0^{+}_1 \to 3^{-}_1)$&=&$
\frac72 e_3^2 \frac{N(N+3)}{N+1}$ \\[5pt]
\hline
$\displaystyle B(E1;3^{-}_1 \to 2^{+}_1)$&=&$\frac{3}{70} 
\left[\alpha_{\nu }N_{\nu }+\alpha_{\pi } N_{\pi }+
\beta \left(N_{\pi }+\eta N_{\nu }\right)\frac{2\,N-1}{2N} \right]^2
\frac{(N+3)(N+4)}{N^2(N+1)}  $ \\[5pt]

$\displaystyle B(E1;2^{+}_{\rm ms} \to 3^{-}_1)$&=&$\frac{3}{35} 
\left[2\left(\alpha_{\pi }- \alpha_{\nu }\right)\frac{N+1}{N+2}+ 
\beta \left(1-\eta \right)\left(1-\frac{N+1}{N(N+2)}\right) \right]^2
\frac{(N+2)(N+3)N_{\nu } \,N_{\pi }}{N^2(N+1)^2}$ \\[5pt]

$\displaystyle B(E1;1^{-}_1 \to 0^{+}_1)$&=&$\frac{1}{10} 
\left[\alpha_{\nu }N_{\nu }+\alpha_{\pi } N_{\pi }+
\beta \left(N_{\pi }+ \eta N_{\nu }\right)\frac{2N+7}{2N}\right]^2
\frac{N-1}{N(N+1)} $ \\[5pt]

$\displaystyle B(E1;1^{-}_1 \to 2^{+}_1)$&=&$\frac{3}{1225} \beta ^2 \chi^2
\left(N_{\pi }+\eta N_{\nu }\right)^2
\frac{(N-1)(N+4)(N+5)^2}{N^4(N+1)}$\\[5pt]  
%%* $\displaystyle B(E1;3^{+}_{\rm ms} \to 3^{-}_1)$&=&$\frac{9}{1400}\beta ^2 
%%\chi^2 (1-\eta )^2 \frac{(N+3)(N+4)^2}{N^3(N+1)^2}  N_{\pi }N_{\nu }$ \\  
\end{tabular} 
\end{table}

\begin{table}
\label{tab:u7}     
\caption{Partial classification scheme for the chain 
U$_{\pi \nu }$(7)$\supset $SO$_{\pi \nu }$(7)$\supset $G$_{2 \pi \nu }$
$\supset $ SO$_{\pi \nu }$(3)}
\begin{tabular}{llll}

U$_{\pi \nu }$(7)&SO$_{\pi \nu }$(7)&G$_{2\pi \nu }$&SO$_{\pi \nu }$(3)\\
\hline                                                       
$[m_1,m_2]$ & $(\omega_1,\omega_2)$ & $(u_1,u_2)$ & $L$ \\
\hline
$[1]$   & $(1,0)$ & $(1,0)$ & 3 \\     
\hline
$[2]$   & $(2,0)$ & $(2,0)$ & 2,4,6 \\ 
        & $(0,0)$ & $(0,0)$ & 0 \\           
$[1,1]$ & $(1,1)$ & $(1,0)$ & 3 \\
        &         & $(1,1)$ & 1,5 \\      
\hline       
$[3]$   & $(3,0)$ & $(3,0)$ & 1,3,4,5,6,7,9 \\ 
        & $(1,0)$ & $(1,0)$ & 3 \\
$[2,1]$ & $(1,0)$ & $(1,0)$ & 3 \\
        & $(2,1)$ & $(2,0)$ & 2,4,6 \\     
        &         & $(2,1)$ & 2,3,4,5,7,8 \\  
        &         & $(1,1)$ & 1,5 \\
\hline
$[4]$   & $(4,0)$ & $(4,0)$ & $0,2,3,4^2,5,6^2,7,8^2,9,10,12$ \\ 
        & $(2,0)$ & $(2,0)$ & 2,4,6 \\
$[3,1]$ & $(2,0)$ & $(2,0)$ & 2,4,6 \\
        & $(3,1)$ & $(3,0)$ & 1,3,4,5,6,7,9 \\     
        &         & $(3,1)$ & $1,2,3^2,4,5^2,6^2,7^2,8,9,10,11$ \\  
        &         & $(2,1)$ & 2,3,4,5,7,8 \\  
        & $(1,1)$ & $(1,0)$ & 3 \\
        &         & $(1,1)$ & 1,5 \\   
$[2,2]$ & $(2,0)$ & $(2,0)$ & 2,4,6 \\ 
        & $(0,0)$ & $(0,0)$ & 0 \\           
        & $(2,2)$ & $(2,0)$ & 2,4,6 \\ 
        &         & $(2,2)$ & 0,2,4,5,6,8,10 \\  
        &         & $(2,1)$ & 2,3,4,5,7,8 \\  
\end{tabular}
\end{table}

\begin{figure}
\label{fig:1-ms}
\epsfig{file=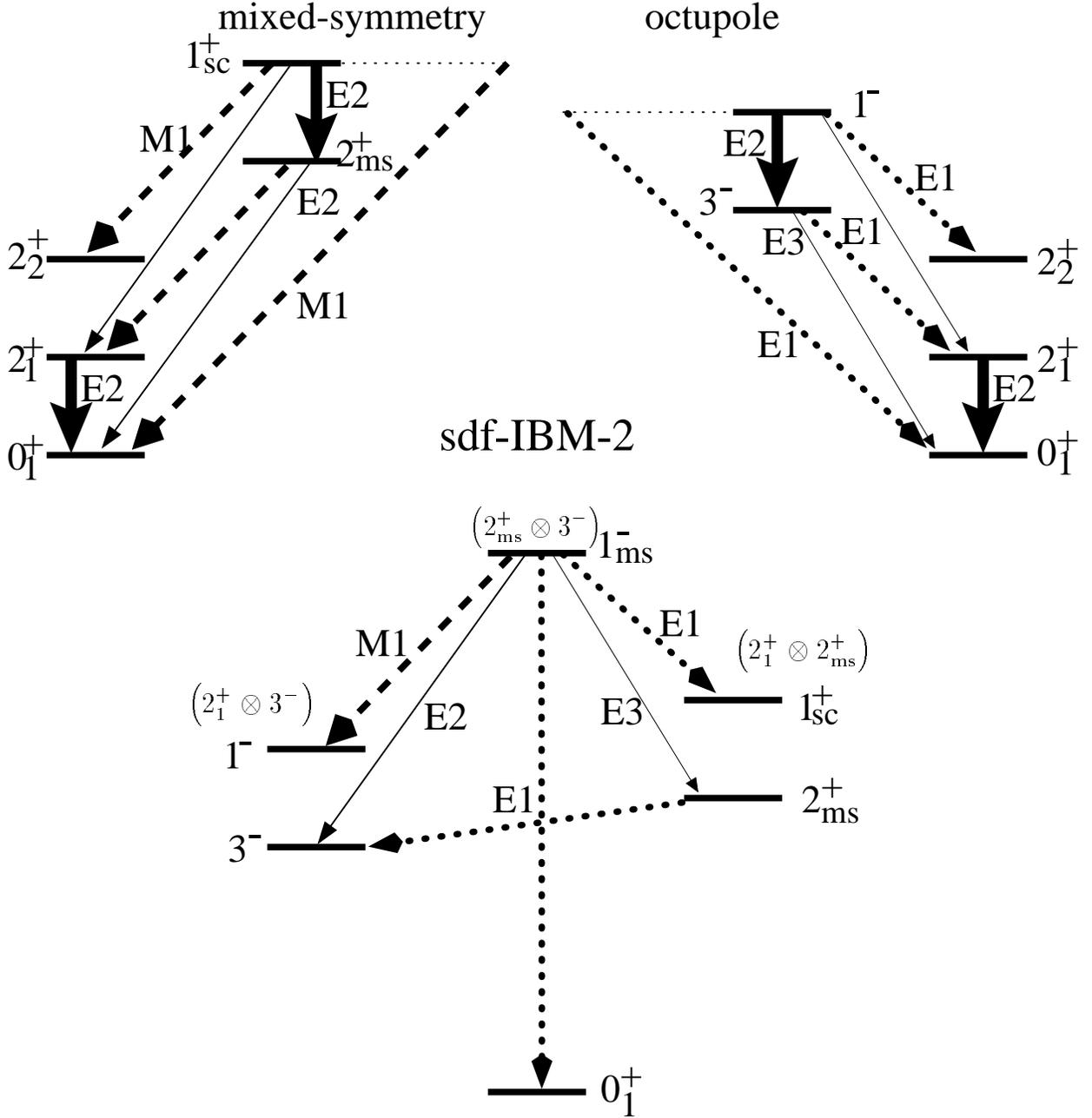,width=\textwidth}
\caption{Expected decays of fundamental one-phonon and 
         two-phonon states in the \protect$sdf$-IBM-2. 
         \protect$2^+_{\rm ms} \to 3^-$ transitions were observed 
         already experimentally \protect\cite{Van95,Franpc}.}
\end{figure}

\begin{figure}
\label{fig:u5}
\epsfig{file=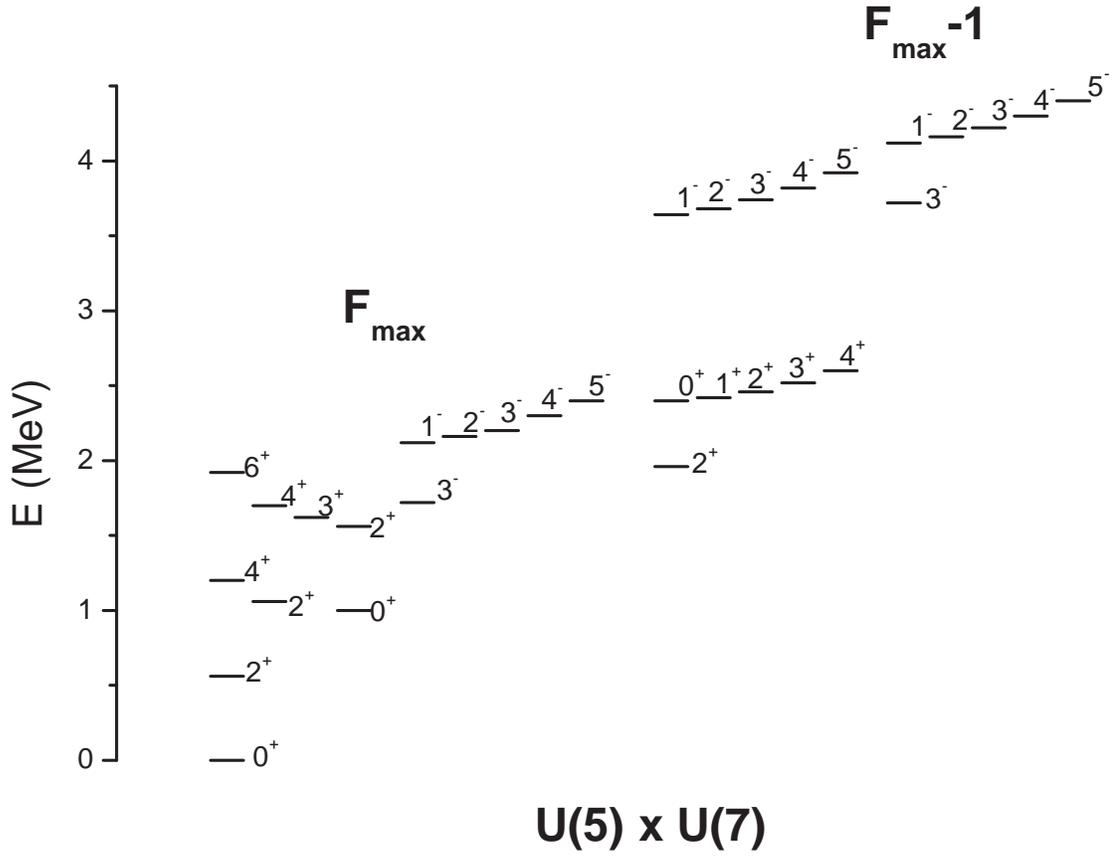,width=\textwidth}
\caption{Sketch of the relevant low-energy spectrum in the U(5)$\otimes $U(7) 
         dynamical symmetry limit calculated for five bosons 
         ($N_{\pi }=4$, $N_{\nu }=1$) with the set of parameters:
         $\epsilon_d=0.5$ MeV, $\epsilon_f=1.6$ MeV, 
         $\gamma=0.01$ MeV, 
         $\lambda =0.4$ MeV and $\lambda' =-0.12$ MeV. 
         The other parameters from Eq. 
         (\protect\ref{HU5}) are put equal to 0.}
\end{figure}

\begin{figure}
\label{fig:so6}
\centerline{\epsfig{file=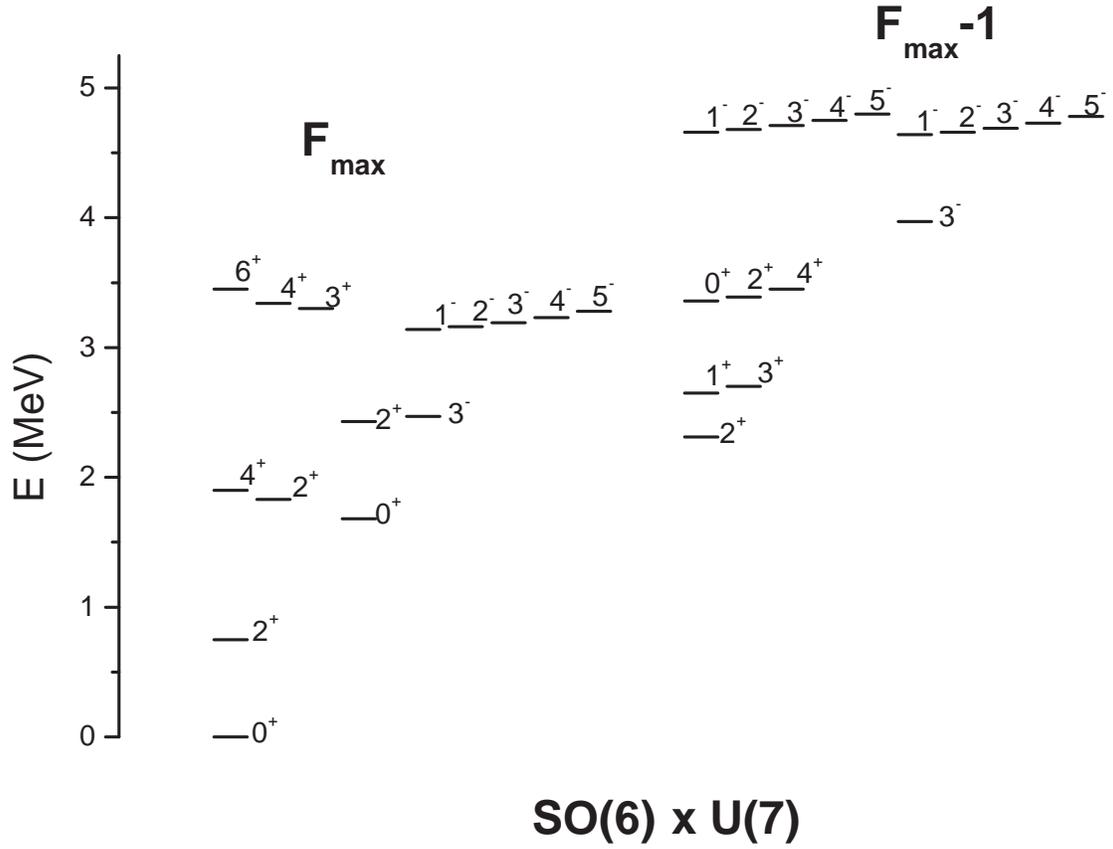,width=\textwidth}}
\caption{Sketch of the relevant low-energy spectrum in the SO(6)$\otimes $U(7) 
dynamical symmetry limit with the set of parameters: $\zeta =-0.07$ MeV,
$\beta =0.18$ MeV, $\epsilon_f=1.5$ MeV, $\gamma=0.005$ MeV,
$\lambda =0.3$ MeV, $\lambda' =-0.1$ MeV and $H_0=3.15$ MeV 
(the other parameters 
from (\protect\ref{HSO6}) are equal to 0) and $N_{\pi }=4$, $N_{\nu }=1$.}
\end{figure}

\end{document}